\pdfoutput=1 
\documentclass[useAMS,usenatbib]{mn2e}

\usepackage{graphicx}
\usepackage{times}
\usepackage{natbib}
\usepackage{color}
\usepackage{ulem}
\usepackage{lipsum}

\usepackage{hyperref}
\hypersetup{
    colorlinks = true,
    linkbordercolor = {1.0 1.0 1.0},
    allcolors = {cyan},
}

\usepackage{amsmath}
\usepackage{amssymb}
\usepackage{float}
\usepackage{todonotes}

\newcommand{\apjs}{ApJS}

\newcommand{\apjl}{ApJL}

\newcommand{\aap}{A \& A}

\begin{document}
 
\title[Morphology of radio relics] {Morphology of radio relics I: What causes the substructure of synchrotron emission?}
\author[P. Dom\'inguez-Fern\'andez et al.]{P. Dom\'inguez-Fern\'andez$^{1}$\thanks{E-mail: pdominguez@hs.uni-hamburg.de   } , M. Br{\"u}ggen$^1$, F. Vazza$^{2,3,1}$ , W.~E.~Banda-Barrag\'an$^1$, 
\newauthor
K. Rajpurohit$^{2,3}$, A. Mignone$^{4}$, D. Mukherjee$^{5}$, B. Vaidya$^{6}$  \\
$^{1}$ Hamburger Sternwarte, Gojenbergsweg 112, 21029 Hamburg, Germany\\
$^2$  Dipartimento di Fisica e Astronomia, Universit\'{a} di Bologna, Via Gobetti 92/3, 40121, Bologna, Italy\\
$^{3}$  Istituto di Radio Astronomia, INAF, Via Gobetti 101, 40121 Bologna, Italy \\
$^{4}$  Dipartimento di Fisica, Universit\`a di Torino,
  via Pietro Giuria 1, 10125 Torino, Italy \\
$^{5}$ Inter-University Centre for Astronomy and Astrophysics, Post Bag 4, Pune - 411007, India \\
$^{6}$ Discipline of Astronomy, Astrophysics and Space Engineering, Indian Institute of Technology Indore, Khandwa Road, Simrol, Indore 453552, India
}

\date{Received / Accepted}
\maketitle

\begin{abstract}
High-resolution radio observations of cluster radio relics often show complex spatial and spectral features. However, it is not clear what these features reveal about the underlying magnetic field properties. We performed three-dimensional magneto-hydrodynamical simulations of merger shock waves propagating through a magnetised, turbulent intracluster medium. Our model includes the diffusive shock acceleration of cosmic-ray electrons, their spatial advection and energy losses at run-time. With this set-up we can investigate the relation between radio substructure and pre-shock plasma conditions in the host cluster. We find that upstream turbulence plays a major role in shaping the properties of radio relics produced downstream. Within the assumption of diffusive shock acceleration, we can reproduce the observed discrepancy between the X-ray derived Mach number of shocks, and the Mach number inferred from radio spectra. Our simulated spectral index maps and profiles across the radio relic also suggest that the standard deviation of the upstream magnetic field must be relatively small ($\sigma_B\leq 1 \, \mu$G) in order to reproduce observations and therefore, radio relics can potentially constrain the distribution of magnetic fields in galaxy clusters outskirts. 

\end{abstract}

\label{firstpage} 

\section{Introduction}
\label{sec:intro}

Radio observations of galaxy clusters reveal non-thermal processes in the intracluster medium (ICM). The size of diffuse radio sources is of the order of Mpc. Such large-scale emission is typically grouped into two  categories: \textit{radio haloes} and \textit{radio relics}. The first category refers to objects located at the cluster center and of a relatively regular and spherical shape, with little to absent polarisation \citep[e.g.][]{2014IJMPD..2330007B}. The second category, refers to objects located at the cluster periphery, with elongated shapes and typically large degrees of polarisation (see \citealt{2012SSRv..166..187B} and \citealt{2019SSRv..215...16V} for reviews). Radio relics are the main focus of this work.

The particle acceleration mechanisms causing diffuse radio sources are not fully understood. However, it seems clear that shocks generated during the assembly of galaxy clusters play a key role in accelerating the synchrotron-emitting cosmic-ray electrons (see \citealt{2011JApA...32..505V} or \citealt{2019SSRv..215...27B} for a review). 
The radio emission observed in relics is compatible with synchrotron emission from cosmic-ray (CR) electrons with Lorentz factors of $\gamma_e \sim  10^3$--$10^5$, which are believed to be accelerated via mild shocks ($\mathcal{M}_{\mathrm{radio}}$\footnote{ $\mathcal{M}_{\mathrm{radio}}$ is the Mach number inferred from radio observations.} $\sim 1.7$--$4.6$) 
crossing the ICM \citep[e.g.][]{2006AJ....131.2900C,2010Sci...330..347V,2012A&A...546A.124V}.

The first-order Fermi acceleration process, commonly known as diffusive shock acceleration (DSA), explains the acceleration of relativistic particles by the passage of a collisionless shock wave \citep[e.g.][]{bo78,1983RPPh...46..973D}. A small fraction ($\ll 10^{-3}$) of thermal particles can cross the shock front multiple times, and receive a boost in their momentum proportional to $\Delta v/c$, where $\Delta v$ is the velocity jump (difference) across the shock. This acceleration mechanism is observed to be much more efficient than what is expected from theory (see \citealt{2019SSRv..215...16V,2020A&A...634A..64B}). Hence, it has been proposed that the electrons are pre-accelerated \citep[e.g.][]{2012ApJ...756...97K,2013MNRAS.435.1061P} before they enter the DSA mechanism.

Moreover, the DSA process does not offer a straightforward explanation for the non-detection of $\gamma$-rays from clusters (see \citealt{2010ApJ...717L..71A,2014ApJ...787...18A,2016ApJ...819..149A} and \citealt{2014MNRAS.437.2291V}).
Among the most relevant open challenges in our understanding of radio relics are: (i) the discrepancy between the Mach numbers detected in X-rays and those inferred from radio observations assuming DSA \citep[e.g.][]{2020A&A...634A..64B}, and (ii) the high electron acceleration efficiency of the order of several percent for the weak shocks commonly associated with radio relics. Up until now, only a few radio relics can readily be explained by the DSA model \citep[e.g.][]{2020MNRAS.tmpL..75L}.

Recent high-resolution radio observations have shown a plethora of complex structures in radio relics \citep[e.g.][]{2018ApJ...852...65R,2020A&A...636A..30R,2014ApJ...794...24O,2017ApJ...835..197V,2018ApJ...865...24D}.
Attempts to explain the observed features struggle with the vast range of scales, from  cosmological scales ($\gtrsim$ Mpc), down to turbulent scales ($\sim 10$ kpc) \citep[e.g.][]{2016arXiv160105083E,2019MNRAS.486..623D}, or even down to  plasma scales where particle acceleration occurs ($\sim 10^{-6} \rm ~kpc$ for the largest gyroradius of relativistic protons). A possible choice is to study particle acceleration from the microphysical point of view using Particle In Cell (PIC) simulations \citep[e.g.][]{2014ApJ...797...47G,2014ApJ...783...91C,2015PhRvL.114h5003P,2014ApJ...794...46C,2019ApJ...883...60R,2019ApJ...876...79K}, or conversely on larger scales using the magneto-hydrodynamical (MHD) approximation, as customarily done with cosmological simulations (the reader may refer to \citealt{review_dynamo} for a review).

 Radio emission from radio relics has been modelled in previous works on larger scales (e.g. \citealt{2013ApJ...765...21S,2015ApJ...812...49H,2017MNRAS.470..240N}). Due to the discrepancy in scales, it is not possible for MHD simulations to model the emission produced by single electrons, but rather from an ensemble of  \textit{tracer particles}, representing a whole distribution of electrons. Previously, this approach has been applied to cosmological MHD simulations in post-processing \citep[see][]{wittor2019}. 
 
In this paper, we model the synchrotron emission at run-time in a small fraction of the ICM by means of a new hybrid particle and fluid framework using the MHD code PLUTO \citep[][]{pluto1,2018ApJ...865..144V}. Our aim is to study the substructure observed in radio relics \citep[e.g.][]{2020A&A...636A..30R}. This method uses \textit{Lagrangian} particles embedded in a large-scale MHD flow, each with its individual energy spectrum.  Here we consider a simplified scenario: we set up a shock tube in a turbulent medium that is representative of a small region of the ICM. We then assume that CR electrons are injected instantly at the shock discontinuity and acquire spectral properties according to DSA theory.

The paper is structured as follows: in Section \ref{section:num_set-up}, we describe our numerical set-up and initial conditions. In Section \ref{section:methods}, we include a description of the particles' initial spectral distribution and evolution. In Section \ref{sec:maps}, we explain how we obtain the emission and spectral index maps. Section \ref{sec:results} shows our results and we summarize in Section \ref{sec:conclusions}.

\section{Numerical set-up}\label{section:num_set-up}

\subsection{Initial conditions: modelling the turbulent ICM with FLASH}
\label{sec:turb}

The turbulent ICM initial conditions were produced using the MHD FLASH code version 4.6.1 \citep{2000ApJS..131..273F, 2002ApJS..143..201C},  with the goal of generating realistic pre-shock conditions. We used the unsplit staggered mesh (USM) MHD solver which uses a constrained transport (CT) method at cell interfaces for preserving the divergence-free magnetic field property on a staggered grid \citep[e.g.][]{2009ASPC..406..243L}. The simulation domain is chosen to be a cubic box of size $L=L_x=L_y=L_z=200$ kpc, uniformly spaced over a $256^3$ cells grid, with periodic boundary conditions. We assumed an ideal gas equation of state with $\gamma_{0}=5/3$.

Turbulence was created following a spectral forcing method, based on the stochastic Ornstein-Uhlenbeck (OU) process to model the turbulent acceleration $\mathbf{f}$ with a finite autocorrelation time \citep[e.g.][]{1988PhFl...31..506E,2006A&A...450..265S,2010A&A...512A..81F}. The OU process describes the evolution of the forcing term in Fourier space, $\hat{\mathbf{f}}$, with a stochastic differential equation:
\begin{equation}
    d \hat{\mathbf{f}}(\mathbf{k},t) = f_0(\mathbf{k}) \mathcal{P}^{\zeta}(\mathbf{k})d\mathbf{W}(t) - \hat{\mathbf{f}}(\mathbf{k},t) \frac{dt}{T},  
\end{equation}
where $f_0$ is the forcing amplitude, $\mathbf{W}(t)$ is a random Wiener process, $\mathcal{P}^{\zeta}$ is a projection tensor in Fourier space, $\zeta$ is the forcing parameter ($\zeta =0(1)$ for purely compressive(solenoidal) forcing), and T is the autocorrelation time scale of the forcing (the reader may refer to \citealt{2010A&A...512A..81F} for a detailed explanation on turbulence forcing). In this work, we solely focus on solenoidal subsonic turbulence forcing ($\nabla \cdot \mathbf{f}=0$), since several authors have shown that the most dominant type of turbulence in the ICM should be subsonic with a large ($\geq 70\%$) predominance of solenoidal modes \citep[e.g.][]{mi14,va17turbo}. 

We have run two simulations with two different stirring scales. The forcing amplitude, $f_0$, was set to be a paraboloid in Fourier space in both simulations only containing power on the largest scales in a small interval of wavenumbers. We chose two different intervals: $1 \leq kL/2\pi \leq 2$ for the first simulation and $1 \leq kL/2\pi \leq 4$ for the second simulation. As the power peaks in 2/3 and 1/4 of the box, we will refer to each of them as $2L/3$ and $L/4$, representing injection scales of $133 \rm ~kpc$ and $50 \rm ~kpc$, respectively.  Furthermore, the maximum $k$ allowed in the box corresponds to the 2L/3 case. Conversely, the $L/4$ case satisfies the minimum condition where its eddy turnover time is smaller than the time needed for our main set-up (where a shock sweeps this turbulent medium). 

\begin{figure}
    \centering
    \includegraphics[width=\columnwidth]{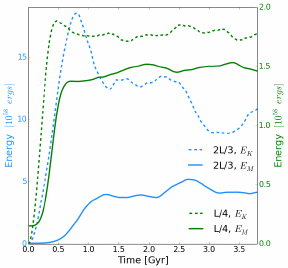}
    \caption{Energy evolution for the runs with injection at $2L/3$ (\textit{blue}) and at $L/4$ (\textit{green}). The kinetic energy is shown with \textit{dashed lines} and the magnetic energy with \textit{solid lines}. Saturation is reached at $t=2t_{2L/3}=$2.1 Gyr and $t=2t_{L/4}=$0.78 Gyr, respectively. After saturation the energy ratio, $E_M/E_K$, is $\sim 0.5$ and $\sim 0.85$ for the 2L/3 and $L/4$ cases, respectively.}
    \label{fig:EkM}
\end{figure}

\begin{figure}
    \centering
    \includegraphics[width=\columnwidth]{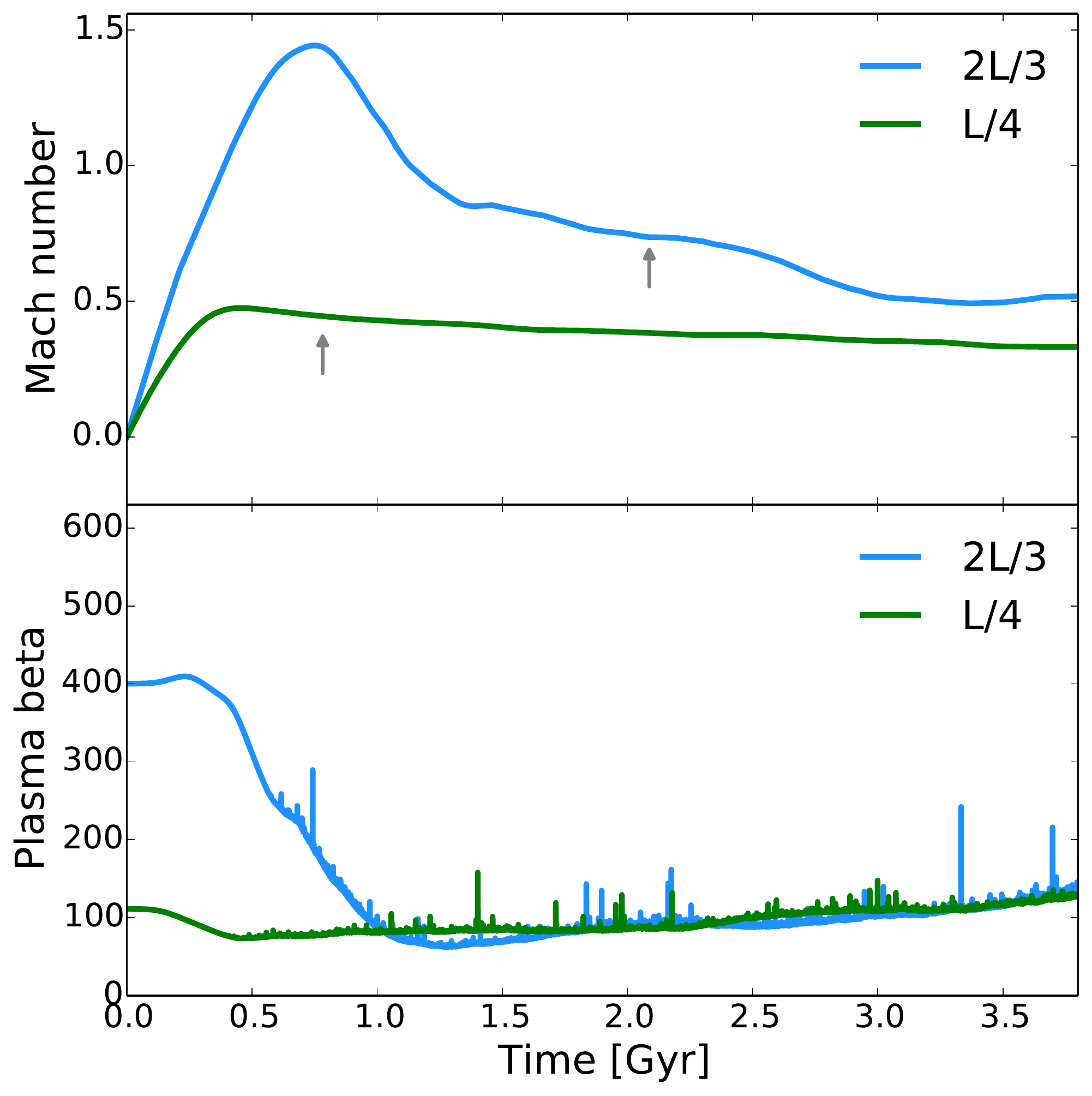}
    \caption{Top panel: Evolution of rms Mach number for the runs with injection at $2L/3$ (\textit{blue}) and at $L/4$ (\textit{green}). The arrows point to the selected snapshots to be our initial conditions. Bottom panel: Corresponding evolution of the plasma beta, $\beta$. The selected snapshots have an rms Mach number of $\mathcal{M}_s \sim$0.45-0.7 and a plasma beta of $\beta \sim$110.}
    \label{fig:Mach_beta}
\end{figure}

The autocorrelation timescale was set equal to the dynamical timescale (also called eddy turn-over timescale) on the scale of energy injection, $t_{2L/3}=2L/3\sigma_v$ and $t_{L/4}=L/4\sigma_v$, respectively, where $\sigma_v$ is the rms velocity to be achieved at saturation. Both simulations were set to have an amplitude of the fluctuations of $\sigma_v=125$ km/s. The turbulence is fully developed after roughly two dynamical timescales when the magnetic energy, the plasma beta, and the rms Mach number become stable (although some transient fluctuations can still be present depending on the balance between mechanical energy from the forcing term and the dissipation rate). At this point, the magnetic field reaches a saturated state since we start with a relatively strong magnetic field seed of 0.2 $\mu$G (for the $2L/3$ case) and 0.4 $\mu$G (for the $L/4$ case). This is shown in Fig.~\ref{fig:EkM} where we plot the evolution of the total kinetic and magnetic energy and in Fig.~\ref{fig:Mach_beta}, where we show the evolution of the rms Mach number and the plasma beta, $\beta$, for both runs. Even after the magnetic saturation, the thermal pressure fluctuations will continue to increase due to turbulent dissipation and, as a consequence, the rms sonic Mach number decreases.

\begin{figure}
    \centering
    \includegraphics[width=\columnwidth]{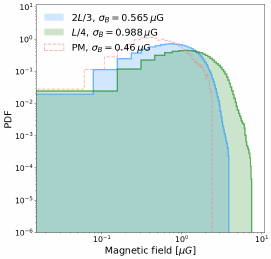}
    \caption{PDFs of magnetic field strength at the selected time for the runs with injection at $2L/3$ (\textit{blue}) and at $L/4$ (\textit{green}). Extra dashed line: PDF of a post-merger (PM) galaxy cluster previously analysed in \citealt{2019MNRAS.486..623D}.}
    \label{fig:PDFs_B}
\end{figure}
We selected one snapshot from each of these two runs to represent a small region of the ICM and each act as an initial condition (see Sec.~\ref{section:num_set-up} and Fig.~\ref{fig:maps_initial}). 
The snapshots are taken at the respective saturation times, which is $t=2.1$ Gyr for the run with injection scale $2L/3$ and $t=0.78$ Gyr for the run with injection scale $L/4$.
In Fig.~\ref{fig:PDFs_B}, we show the probability distribution function of the magnetic field strength at those times. The two snapshots have a sufficiently different distribution of magnetic fields (due to the difference in the initial magnetic field seed and the interval of wavenumbers of the stirring), yet in terms of the rms sonic Mach number and plasma-$\beta$ they lie in the range of characteristic values of the ICM \citep[e.g.][]{ry08}. Note that the tail of the $L/4$ distribution extends to $\sim 2$ times larger magnetic field values. (see Fig.~\ref{fig:PDFs_B}). For reference we include the PDFs of one galaxy cluster previously produced in a cosmological MHD simulation and analysed in \citealt{2018MNRAS.474.1672V,2019MNRAS.486..623D}. This cluster is classified as a post-merger (PM) cluster with a mass $\mathrm{M}_{200}=1.9 \times 10^{15}\mathrm{M}_{\odot}$ \footnote{The reader can find this galaxy cluster in \citealt{2019MNRAS.486..623D} with ID E16B}. We selected a region at a distance of $\sim 1$ Mpc from the cluster's center with an extent of $\sim 250$ kpc. As can be seen in Fig.~\ref{fig:PDFs_B}, the PDF of the magnetic field broadly agrees with previous results from cosmological MHD simulations. The high-magnetic field tail of the distribution is slightly more extended than in cosmological simulations, owing to the larger Reynolds number in these new simulations.  The magnetic field strength at the outskirts of the
clusters is most likely underestimated in the cosmological simulation due to the limited resolution \citep[see][]{2018MNRAS.474.1672V}. 
A more extensive survey of the interaction between merger shocks with a larger range of different initial turbulent conditions for the ICM will be the subject of future work.

\subsection{Main PLUTO simulations}
\label{sec:PLUTO}

In order to study the synchrotron emission in an MHD shock tube, we use the code PLUTO \citep{pluto1}, which solves the following conservation laws for ideal MHD:
\begin{small}
\begin{align}
 \frac{\partial \rho}{\partial t} + \mathbf{\nabla} \cdot (\rho \mathbf{v}) = &0,  \\
 \frac{\partial \mathbf{m}}{\partial t} + \mathbf{\nabla} \cdot \left[ \mathbf{m} \mathbf{v} - \mathbf{B}\mathbf{B} + \mathbf{I} \left( p + \frac{B^2}{2} \right) \right]^T = 0,  \\
 \frac{\partial (E_t)}{\partial t} + \mathbf{\nabla} \cdot \left[ \left( \frac{\rho v^2}{2} + \rho e + p  \right)\mathbf{v} - (\mathbf{v} \times \mathbf{B}) \times \mathbf{B} \right] = & 0,  \\ 
 \frac{\partial \mathbf{B}}{\partial t} - \mathbf{\nabla} \times (\mathbf{v} \times \mathbf{B}) = &0,  \\
 \mathbf{\nabla} \cdot \mathbf{B} = &0,  \\
 \rho e =  \frac{p}{\gamma_0 - 1}& , 
\end{align}
\end{small}
where $\rho$ is the gas mass density, $\mathbf{m}=\rho \mathbf{v}$ is the momentum density, $p$ is the thermal pressure, $\mathbf{B}$ is the magnetic field (where a factor of $1/\sqrt{4\pi}$ has been absorbed in its definition), $E_t$ is the total energy density, $e$ the specific internal energy and where we assumed an ideal equation of state (EOS), that is $\gamma_{0} = 5/3$.

\begin{figure}
    \centering
    \includegraphics[width=\columnwidth]{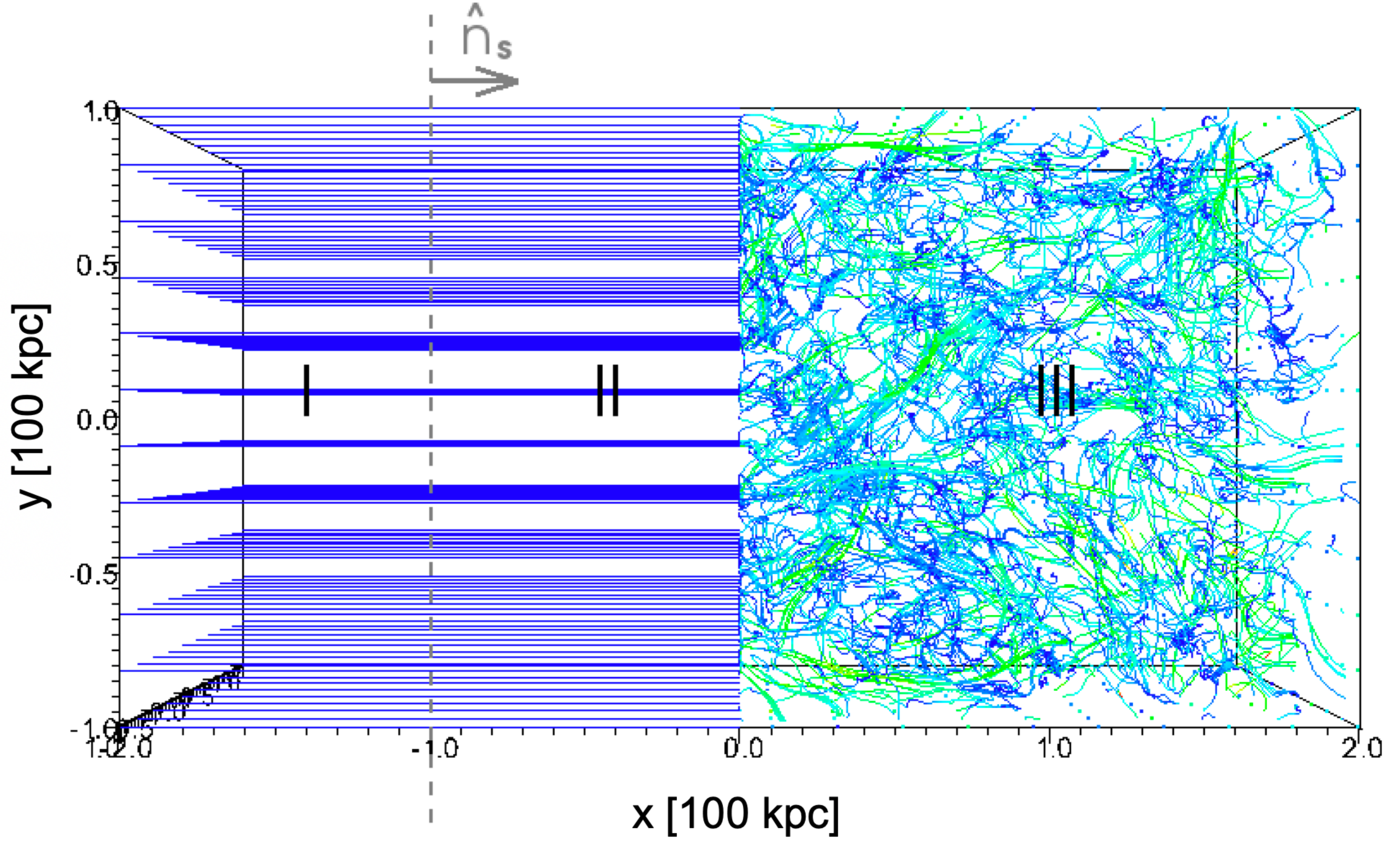}
    \caption{Initial magnetic field configuration in the PLUTO code. The streamlines are coloured according to the magnitude of the magnetic field: green, light colours denote higher values, while dark blue colour indicates lower values. I denotes the post-shock region, II denotes the uniform pre-shock region and III the turbulent pre-shock region  (see Sec.~\ref{sec:turb}). The left side is a uniform medium with a $B_x$ component matching the mean value of the $B_x$ of the turbulent medium. We have one Lagrangian particle per cell placed in the whole regions II and III.}
    \label{fig:init}
\end{figure}

Our computational domain is a rectangular box (400 kpc $\times$ 200 kpc $\times$ 200 kpc with $256 \times 128 \times 128$ cells, respectively), where $x\in$[-200,200] kpc, $y \in$[-100,100] kpc, and $z\in$[-100,100] kpc (see Fig.~\ref{fig:init}). The right-hand half of the domain is filled with a turbulent medium
(see Sec.~\ref{sec:turb}), representing
a realistic ICM, while the left-hand half contains a uniform medium in which the shock is launched. We define a shock discontinuity at $x=-100$ kpc (see Fig.~\ref{fig:init} for the initial configuration of the magnetic field). This defines three regions in our simulation box: a post-shock uniform region (I), a pre-shock uniform region (II) and a pre-shock turbulent region (III).

\begin{figure}
    \centering
    \includegraphics[width=\columnwidth]{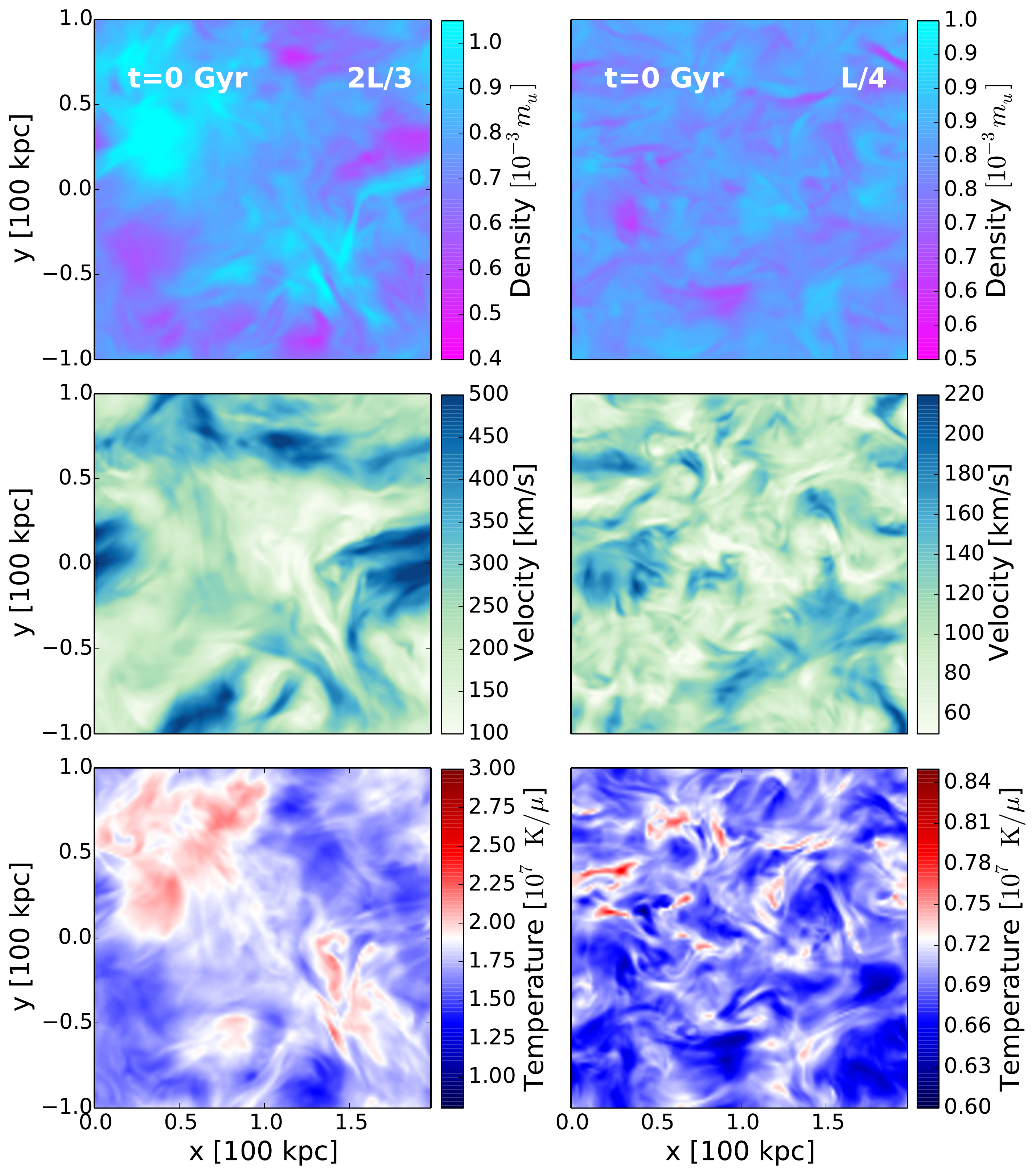}
    \caption{Projected maps along the $z$-axis of the different initial conditions (25 kpc slice), where $m_u$ is the atomic mass unit and $\mu$ is the mean molecular weight. Here we show the right-hand half-size of the box containing the turbulent media. From top to bottom the reader can see the average density, velocity and temperature field for the $2L/3$ (\textit{left column}) and $L/4$ (\textit{right column}) cases, where $2L/3$ and $L/4$ refer to the integral scale of each type of turbulence ($L =200$ kpc).}
    \label{fig:maps_initial}
\end{figure}

The turbulent medium is produced externally, with the procedure outlined in Sec.~\ref{sec:turb}. The turbulent fields are then read into PLUTO and interpolated onto the numerical mesh used to evolve shocks with a  bi- or tri-linear interpolation at the desired coordinate location at the initial time.
The initial boundary conditions of the computational domain are \textit{outflow} in $x$ (zero gradient across the boundary) and \textit{periodic} in $y$ and $z$. We used a piecewise parabolic method (PPM) for the spatial integration, whereas a $2^{nd}$ oder TVD Runge-Kutta method for the time stepping with a Courant-Friedichs-Lewy (CFL) condition of $0.3$. The Riemann solver for the flux computation that we used is the Harten-Lax-van Leer-Discontinuities (HLLD) solver \citep[see][]{2005JCoPh.208..315M}. We control the $\nabla \cdot \mathbf{B} = 0$ condition with the hyperbolic divergence cleaning technique where the induction equation is coupled to a generalized Lagrange multiplier (GLM) \citep[e.g.][]{2002JCoPh.175..645D}. In Appendix \ref{appen_2}, we compare how the GLM and constrained transport (CT) methods control the divergence-free condition and find that both methods work reliably in PLUTO. \\
The initial conditions for the density, pressure and velocity in region II (\textit{pre-shock} uniform region at [-100,0] kpc) 
 are set to the mean value of the corresponding turbulent fields. In the case of the magnetic field in region II, we set it to be the mean value of the $B_x$ component of the turbulent medium. The initial conditions for region I (\textit{post-shock} region) are selected according to the MHD Rankine-Hugoniot conditions \citep[e.g.][]{Landau1987Fluid}. We performed simulations with two different turbulent media (see Sec.
~\ref{sec:turb}) produced with the code FLASH \citep[e.g.][]{2000ApJS..131..273F, 2002ApJS..143..201C} by varying also the strength of the shock and the angle, $\theta_{bn}$, of the upstream magnetic field\footnote{Note that here we define the direction of the upstream magnetic field as the direction of the mean magnetic field of the turbulent medium} with respect to the normal of the shock. Shocks can be classified as quasi-parallel and quasi-perpendicular if $\theta_{bn} \leq 45^{\circ}$ or $\theta_{bn} > 45^{\circ}$, respectively. We consider two limits, i.e. $\theta_{bn} = 0^{\circ}$ and $\theta_{bn} = 90^{\circ}$. 
This sums up to a total of six runs for which all the parameters are summarized in Table \ref{table:init}. 
We show the projection maps of the two different initial turbulent media considered for this work in Fig.~\ref{fig:maps_initial} and clarify how these were selected in Sec.~\ref{sec:turb}.
\\
Finally, we fill the computational domain from the shock discontinuity up to the right side of the box with one \textit{Lagrangian} particle per cell. This gives us a total number of 3,145,728 Lagrangian particles for each run. 

\begin{table*}
\centering
\begin{tabular}{cccccc}
    &&& \\ \hline
     Run ID & Turbulent medium & $\mathcal{M}$ &$\theta_{bn}[^{\circ}]$ &$\rho_{\mathrm{II}} \, [10^{-27}\mathrm{g/cm}^3] $ & 
       $B_{x,\mathrm{II}} \, [\mu G]$ \\ \hline
      k1p5\_M2\_parallel & $2L/3$ & 2.0 & 0 & 1.338 & 0.4\\ 
      k1p5\_M3\_parallel & $2L/3$ & 3.0  & 0 & 1.338 & 0.4 \\ 
      k4\_M2\_parallel & $L/4$ & 2.0 & 0 & 1.338 & 0.4 \\
      k4\_M3\_parallel & $L/4$ & 3.0 & 0 & 1.338 & 0.4 \\ 
      k1p5\_M3\_perpendicular & $2L/3$ & 3.0 & 90 & 1.338 & 1.156$\times 10^{-12}$\\
      k4\_M3\_perpendicular & $L/4$ & 3.0 & 90 & 1.338& 1.156$\times 10^{-12}$ \\
      \hline
\end{tabular}
\caption{Initial conditions: We denote our regions I, II, III where I is the post-shock region ([-2,-1] in box coordinates), II is the uniform pre-shock region ([-1,0] in box coordinates) and III is the turbulent pre-shock region ([0,2]). The initial conditions for the left side of the shock (region I) depend on the pre-shock conditions (region II) and the initial Mach number of the shock $\mathcal{M}$ through the Rankine-Hugoniot jump conditions. $L$ denotes the length of the turbulent region, i.e. 200 kpc (see more details in Sec.~\ref{sec:turb}). Note that the magnetic field in region II has only an x-component.}
\label{table:init}
\end{table*}

\section{Non-thermal radio emission from shocks}\label{section:methods}

\subsection{Particle energy spectrum}
\label{sec:spectrum}

Each Lagrangian particle represents an ensemble of CR electrons and is characterized by an energy distribution function,
\begin{equation}
    \chi(E,\tau) = \frac{N(E,\tau)}{n_0},
\end{equation}
which gives the number of electrons per fluid number density. These particles evolve according to the cosmic-ray transport equation,
\begin{equation}\label{cr_eq}
    \frac{d\chi}{d\tau} + \frac{\partial}{\partial E} \left[ \left(   - \frac{E}{3}\nabla_{\mu}u^{\mu} - c_rE^2 \right) \chi \right]=0,
\end{equation}
where the first term in square brackets accounts for energy losses due to adiabatic expansion, while the second one accounts for the synchrotron and inverse-Compton losses for electrons with isotropically distributed velocity vectors,
\begin{equation}\label{eq:cr}
    c_r = \frac{4}{3} \frac{\sigma_Tc\beta^2}{m_{e}^{2}c^{4}} \left[ \frac{B^2}{8\pi} + a_{\rm rad}T_{\mathrm{CMB}}^4(1+z)^4
    \right] ,
\end{equation}
where $\beta=v_e/c$ is the velocity of the electrons, $m_e$ their mass and $a_{\rm rad}$ the radiation constant. For the present work, we assume a constant redshift of $z=0$. The reader may refer to \citet{2018ApJ...865..144V} for a complete description of the numerical implementation and the semi-analytical scheme used for solving Eq. (\ref{cr_eq}).

We study a simplified scenario where the non-thermal spectra of the particles are activated instantly at the shock discontinuity. While the particles follow the fluid flow since $t=0$, the particle's energy distribution starts to evolve only when the particles have passed a shock. We implemented a shock finder based on converging flows and pressure jumps that we describe in Appendix \ref{appen_1}. 

Once the Lagrangian particles are \textit{activated} at the shock discontinuity, they get assigned an initial energy distribution. The corresponding particle momentum distribution function is a power-law distribution, i.e $f(p) \propto p^{-q}$. The power-law index is given by the diffusive shock acceleration (DSA) theory \citep{1983RPPh...46..973D}. 
\begin{equation}\label{eq:Drury}
    q = \frac{3r}{r-1},
\end{equation}
where $r$ is the shock compression ratio defined as the ratio of the upstream and downstream densities. If we consider test particle acceleration at a shock of Mach number 
$\mathcal{M}$, then it is possible to re-write Eq. ~(\ref{eq:Drury}) making use of the Rankine-Hugoniot jump equations (see derivation in \citealt{1987PhR...154....1B}):
\begin{equation}\label{spectral_index}
    q = \frac{3(\gamma_0 + 1)\mathcal{M}^2}{2(\mathcal{M}^2-1)} = \frac{4\mathcal{M}^2}{\mathcal{M}^2-1},
\end{equation}
where for the second equality we have considered an adiabatic index of $\gamma_0=5/3$. 
The corresponding power-law index for the macro-particle distribution function can be also obtained by assuming isotropy, i.e 
$N(p,\mathbf{\tau}) = \int \, \Omega_{\tau} p^2 f \approx 4\pi p^2 f$, where $d \Omega_{\tau}$ is the solid angle around the direction $\mathbf{\tau}$. Since the particles are relativistic, we have that $N(E,\mathbf{\tau})dE = N(p,\mathbf{\tau})dp$. Moreover, from the isotropic condition we also have that $N(E) = 4\pi N(E,\mathbf{\tau})$.

Therefore, the tracer particle energy distribution function at the activation time is given by
\begin{equation}\label{eq:energy_dist}
\chi(E) = \frac{N(E)}{n_0} = \frac{N_0}{n_0}\, E^{-p},     
\end{equation}
where $p=q-2$ is what it is usually called the \textit{injection spectral index}, $N_0$ is the normalisation factor and $n_0$ is the fluid number density. $N_0$ is assigned according to the energy contained in the shock. That is, we considered that the total energy per fluid number density is
\begin{equation}
    \int \chi(E)\, E \, dE = \frac{E_{tot}}{n_0},
\end{equation}
where $E_{\rm tot} = \eta \, E_{\rm shock} = \eta \, \rho_{\rm post} \,v_{\rm shock}^2$ and $\eta$ is the acceleration efficiency. For which finally one can obtain the normalisation factor:
\begin{equation}\label{eq:norm}
    N_0 = \left\{
	     \begin{array}{ll}
		 \frac{\eta \, E_{\rm shock}\,(4-q)}{\left[{E_{\rm max}}^{4-q}\, - \, {E_{\rm min}}^{4-q} \right]}       & \mathrm{if\ } q \ne 4 \\
		 & \\
		 \eta \, E_{\rm shock} \log \left( \frac{E_{\rm max}}{E_{\rm min}}\right) & \mathrm{if\ } q=4 \\
		 
	       \end{array}
	     \right.
\end{equation}
The PLUTO code allows us to compute the maximum and minimum energy at each time step. The maximum energy $E_{\rm max}$ is imposed considering the maximum allowed Larmor radius for each particle. The minimum energy $E_{\rm min}$ is estimated balancing the synchrotron and acceleration time scales (see full explanation in \citealt{2018ApJ...865..144V,2010ApJ...711..445B,2012MNRAS.421.2635M}) so it depends on the acceleration efficiency. Moreover, the acceleration efficiency in collisionless relativistic shocks can also depend on the energy of the particles \citep[see][]{2013ApJ...771...54S}. In this work, for simplicity we assume a fixed acceleration efficiency of $\eta=10^{-3}$ and fixed energy limits of $\gamma_{\rm min}=1$ and $\gamma_{\rm max}=10^5$. This acceleration efficiency agrees with the expectations of DSA for strong shocks \citep[e.g.][]{ka12} and lies in the range of values required to explain observations of radio relics (see \citealt{2020A&A...634A..64B}). Nevertheless, this means that the final synchrotron emission obtained can be re-scaled to the desired $\eta$ since in our case the energy limits remain constant.

\subsection{Synchrotron emission}
\label{sec:sync_emiss}

The synchrotron emission of a tracer particle in a local magnetic field $\mathbf{B}'$ in the direction $\mathbf{\hat n_{los}}'$, per unit solid angle, volume and frequency is given by
\begin{equation}
    \mathcal{J}_{\rm syn}^{'}(\nu_{\rm obs}',\mathbf{\hat n_{los}'},\mathbf{B}')
    = \int N(E')
    \mathcal{P}(\nu_{\rm obs}',E',\phi') dE' d\Omega',
\end{equation}
where $\mathcal{P}(\nu_{\rm obs}',E',\phi')$ is the synchrotron power per unit frequency and unit solid angle emitted by a single particle that has energy $E'$ and $\phi'$ is the angle that the velocity vector of the particle makes with the direction $\mathbf{\hat n_{los}'}$. Following \citealt{1965ARA&A...3..297G} we get
\begin{equation}\label{eq:17}
     \mathcal{J}_{\rm syn}^{'}(\nu_{\rm obs}',\mathbf{\hat n_{los}^{'}},\mathbf{B}')
     = \frac{\sqrt{3} e^3}{4\pi m_e c^2} 
     | \mathbf{B}' \times \mathbf{\hat n_{los}}' | \int N(E') F(\xi) \, dE',
\end{equation}
where $\mathbf{\hat n_{los}}'$ is the unit vector in the direction of the line of sight in the comoving frame and $F(\xi)$ is a Bessel function integral given by
\begin{equation}\label{eq:Bessel}
    F(\xi) = \xi \int_{\xi}^{\infty} K_{5/3}(z')\,dz',
\end{equation}
where
\begin{equation}
\label{J_pol}
    \xi = \frac{\nu_{\rm obs}'}{\nu_c^{'}} = \frac{4\pi m_{e}^3c^5 \nu_{\rm obs}'}{3eE'^2| \mathbf{B}' \times \mathbf{\hat n_{los}'}|},
\end{equation}
where $\nu_{c}'$ is defined as the critical frequency at which the emission peaks. Note that only those particles with a pitch angle coinciding with the angle between $\mathbf{B}'$ and $\mathbf{\hat n_{los}'}$ contribute to the emission along the line of sight in Eq. (\ref{eq:17}).

The emissivity in Eq. (\ref{eq:17}) is measured in the local comoving frame with the emitting volume. If we want the emissivity in a fixed observer's frame then we have to apply a transformation\footnote{The reader should note that the primed quantities in Eq. (\ref{J_syn}) refer to the comoving frame, whereas standard notation refers to the observer's frame.}:
\begin{equation}\label{J_syn}
    \mathcal{J}_{\rm syn}(\nu_{\rm obs},\mathbf{\hat n_{los}},\mathbf{B}) = \mathcal{D}^2
    \mathcal{J}_{\rm syn}^{'}(\nu_{\rm obs}',\mathbf{\hat n_{los}}',\mathbf{B}'), 
\end{equation}
where $\mathcal{D}$ is a Doppler factor given by
\begin{equation}
    \mathcal{D} = \frac{1}{\gamma(1- \mathbf{\hat n_{los}}\cdot \mathbf{v})},
\end{equation}
where $\gamma$ is the Lorentz factor of the tracer particle. 
The unit vectors on the direction of the line of sight in the comoving and observer's frame are related through
\begin{equation}
    \mathbf{\hat n_{los}'} = \mathcal{D} \left[ 
    \mathbf{\hat n_{los}} + \left( \frac{\gamma^2}{\gamma +1} \mathbf{v} \cdot \mathbf{\hat n_{los}} - \gamma
    \right) \mathbf{v}
    \right].
\end{equation}

\subsection{Radio emission maps}
\label{sec:maps}

In the preceding equations, the vector $\mathbf{\hat n_{los}}$ can be selected according to an observing angle $\theta_{\rm obs}$ with respect to the line-of-sight. In this work, we only show results considering $\theta_{\rm obs}=0^{\circ}$, that is we consider an observer's reference frame in which $z$ lies along the line of sight $\mathbf{\hat n_{los}}$ and $x$ and $y$ are in the plane of the sky. The specific intensity (or surface brightness) maps can then be obtained by integrating along a line of sight as
\begin{equation}\label{eq:intensity}
    I_{\nu} = \int \mathcal{J}_{\rm syn}(\nu_{\rm obs},x,y,z)dz,
\end{equation}
in units of [erg cm$^{-2}$ s$^{-1}$ Hz$^{-1}$ str$^{-1}$]. This is doable due to the fact that the emissivity information $\mathcal{J}_{\rm syn}$ of the Lagrangian tracer particles is interpolated back onto the Eulerian grid. 
It is also possible to obtain spectral index maps by means of 
\begin{equation}\label{eq:spectral_index}
    - \alpha(x,y) =
    \frac{\log \left[I_{\nu_{2}}(x,y)/I_{\nu_{1}}(x,y) \right]}{\log(\nu_{2} / \nu_{1})},
\end{equation}
and the integrated spectra (or net flux) can be obtained by integrating the specific intensity $I_{\nu}$ over the area covered by the source in the plane of the sky, that is
\begin{equation}\label{eq:flux}
   F_{\nu}(\nu) = \int I_{\nu}(\nu, x, y)dxdy, 
\end{equation}
in units of [erg s$^{-1}$ Hz$^{-1}$ str$^{-1}$].

\section{Results}
\label{sec:results}

\subsection{Fluid properties}
\label{sec:fluid}

\begin{figure*}
    \centering
    \includegraphics[width=0.9\textwidth]{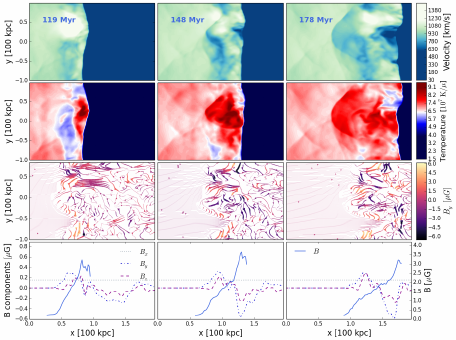}
    \caption{Evolution of the $2L/3$, $\mathcal{M}=3$ and $\theta_{bn}=90^{\circ}$ case. \textit{First row}: 20 kpc slice along the z-axis of velocity field. \textit{Second row}: 20 kpc slice along the z-axis of temperature \textit{Third row}: streamlines of total magnetic field (note that the colour-code only denotes the strength of $y$-component). \textit{Fourth row}: 1D magnetic profiles corresponding. The component's profiles have no weighting, whereas the magnetic field strength, $B$, is weighted with the 150 MHz synchrotron emissivity. }
    \label{fig:evolution}
\end{figure*}
\begin{figure*}
    \centering
    \includegraphics[width=0.68\columnwidth]{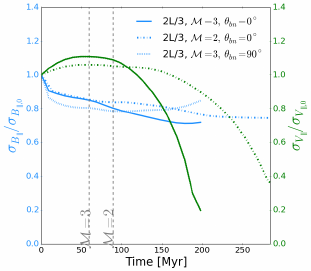}
    \includegraphics[width=0.68\columnwidth]{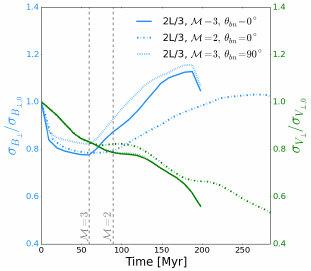}
    \includegraphics[width=0.68\columnwidth]{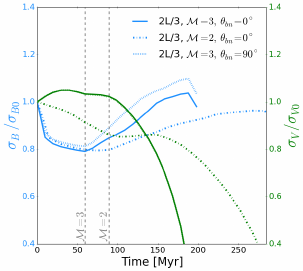}\\
    \includegraphics[width=0.68\columnwidth]{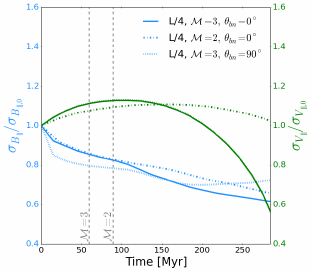}
    \includegraphics[width=0.68\columnwidth]{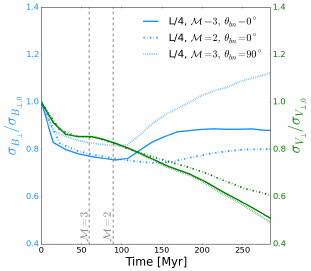}
    \includegraphics[width=0.68\columnwidth]{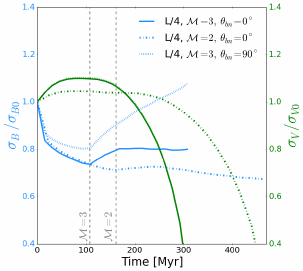}
    \caption{Evolution of the standard deviation of the magnetic field (\textit{left axis}) and velocity field (\textit{right axis}). \textit{Top panels}: $2L/3$ case. \textit{Bottom panels}: $L/4$ case. \textit{First column}: Parallel component to the shock, i.e $B_{\parallel}=B_x$ and $v_{\parallel}=v_x$. \textit{Second column}: Perpendicular component, i.e $B_{\perp}=\sqrt{B_{y}^2 + B_{z}^2}$ and $v_{\perp}=\sqrt{v_{y}^2 + v_{z}^2}$. \textit{Third column}: Standard deviation of the magnetic and velocity field strengths. $\sigma_B$ and $\sigma_V$ are normalised to their value at $t=0$ for purposes of comparison between different runs (see Table \ref{table:std}). }
    \label{fig:std}
\end{figure*}
In this subsection we describe some of the features of the fluid starting with the evolution of velocity, temperature and magnetic field, for the $2L/3$, $\mathcal{M}=3$ and $\theta_{bn}=90^{\circ}$ case (see Fig.~\ref{fig:evolution}). In all of our runs, the initial shock sweeps region II maintaining a constant velocity and planar shape due to the fact that this region is initially uniform. Next, the shock enters region III, where the shock will no longer be uniform and be affected by the anisotropies of the fluid. 
The two first rows in Fig.~\ref{fig:evolution} highlight how weak internal shocks and turbulence are generated in the downstream as the main shock travels through the simulation box. The third row shows the streamlines of the magnetic field. The streamlines are coloured according to the magnitude of the $y$-component of the magnetic field, $B_y$, to illustrate how the shock compression amplifies the field. The fourth row shows 1D profiles of the magnetic field obtained by integrating along the LOS. As expected from the MHD Rankine-Hugoniot conditions, the component parallel to the shock normal, $B_x$, is conserved, while the other components are amplified and stretched as a result of shock compression. We show also the magnetic field profile weighted with the synchrotron emission at 150 MHz (blue solid lines). 

In Fig.~\ref{fig:std} we show how the standard deviation (volumetric value) of the magnetic and velocity field evolve for all runs listed in Tab.~\ref{table:init}. The evolution is characterized by two phases: 1) a first phase in which the shock is crossing region II (lasting 50--150 Myr depending on the type of run), and we have a purely decaying turbulence system on the right-hand side of the box ([0,200] kpc, see Fig.~\ref{fig:init}), and 2) a second phase in which the shock has already entered region III and is compressing the turbulent medium. The dashed gray vertical lines in Fig.~\ref{fig:std} define the beginning of this second phase (the time differs according to the initial Mach number of each run). 

\begin{table}
\centering
\begin{tabular}{cccc}
    \hline
    Run ID & $\sigma_{V0}\mathrm{[km/s]}$ & $\sigma_{B0}[\mu$G] & $t_{\rm shock}$[Myr]   \\ \hline
      k1p5\_M2\_parallel & 217.7 & 1.016 & 282 \\  
      k1p5\_M3\_parallel & 388.5 & 1.016 &188 \\  
      k4\_M2\_parallel & 133.1 & 0.659 & 439 \\ 
      k4\_M3\_parallel & 246.1 & 0.513 & 292 \\  
      k1p5\_M3\_perpendicular & 388.5 & 1.099 & 188 \\
      k4\_M3\_perpendicular & 246.1 & 0.513 & 293 \\ 
      \hline
\end{tabular}
\caption{Values used in Fig. \ref{fig:std} and \ref{fig:mag_profile}. The second and third column are the initial standard deviation of the velocity and magnetic field in the whole simulation box. The fourth column shows the total time needed for the shock to cross the entire simulation box.}
\label{table:std}
\end{table}

The evolution of the standard deviation of the velocity and magnetic field are in general not correlated
(see third column of Fig.~\ref{fig:std}). We also analysed the standard deviation of the parallel and perpendicular components of both fields with respect to the shock normal. In general, we find that the evolution of the parallel component dominates the evolution of $\sigma_{V}$, whereas the perpendicular component dominates the evolution of $\sigma_{B}$. 
The standard deviation of the parallel component (see first column of Fig.~\ref{fig:std}) decreases with time for both fields. The standard deviations of the perpendicular components (see second column of Fig.~\ref{fig:std}) show a different evolution: the perpendicular component, $\sigma_{V_{\perp}}$, follows the same behaviour of $\sigma_{V}$, while the perpendicular $\sigma_{B}$ always increases right after the shock crossing in all  runs. This trend is persistent and gets stronger with higher resolution (see Fig.~\ref{fig:resolution1} in Appendix \ref{appen_3}).

The reason for this is that most velocity fluctuations are driven in the direction parallel to the shock normal, while magnetic field fluctuations are initially driven perpendicular to the shock normal due to compression. 
With time, the dynamics gets more complicated due to the shock compression and possibly also stretching of the magnetic field. 
The increase in $\sigma_{B}$ induces a delay in the velocity field dissipation. The plateau observed in $\sigma_{V}$ in the second phase (evolution to the right of the dashed gray lines in Fig.~\ref{fig:std}) indicates that the shock-induced turbulence could be maintained only for some time before $\sigma_{V}$ decreases due to turbulent dissipation.

For all our runs, we verified that the density and also the temperature evolve in the same fashion as the velocity field in Fig.~\ref{fig:std} and it is only
the standard deviation of the magnetic field that has a characteristic evolution in both phases. 

\begin{figure}
    \centering
    \includegraphics[width=\columnwidth]{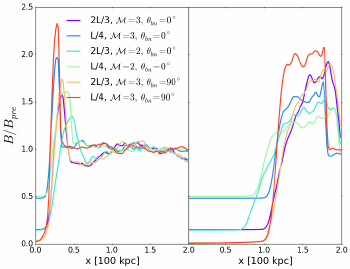}
    \caption{1D magnetic field profile of all runs. \textit{Left panel}: Shock front just entering the turbulent medium at $t_{\rm shock}/9$. \textit{Right panel}: Shock front is almost at the right end of the $x$-axis at $t_{\rm shock}/19$. Refer to Tab.~\ref{table:std} for the value of $t_{\rm shock}$ of each run. }
    \label{fig:mag_profile}
\end{figure}

In the following, we summarize the observed effects for this second phase: 

\begin{enumerate}
    \item[i)] \textit{The role of different turbulent injection scales}: We find that $\sigma_B$ decreases faster in a system with smaller magnetic fluctuations ($L/4$) than in one with larger fluctuations ($2L/3$). This is expected as the turbulence injection scale is smaller in the $L/4$ case and therefore the eddy-turnover time is shorter. In fact, in the $L/4$ case, the shock-induced turbulence seems to only have an increasing effect on $\sigma_B$ whenever the shock is perpendicular, i.e. $\theta_{bn}=90^{\circ}$.
    \item[ii)] \textit{The role of the main shock's Mach number}: The largest impact on increasing $\sigma_B$ is due to a larger Mach number. For example, the $\mathcal{M}=3$ and $\theta_{bn}=0^{\circ}$ case shows an increase of $\sim 31$\% for the $2L/3$ turbulence and $\sim 7$\% for the $L/4$ turbulence. Conversely, the $\mathcal{M}=2$ and $\theta_{bn}=0^{\circ}$ case shows an increase of $\sim 25$\% for the $2L/3$ turbulence and a decrease of $\sim 4$\% for the $L/4$ turbulence. This suggests that weak shocks ($\mathcal{M}=$2) are less likely to modify the initial distribution of magnetic fields for smaller turbulent scales.
    \item[iii)] \textit{The role of obliquity}: A perpendicular shock has the strongest effect on broadening the downstream magnetic field distribution. This is expected as the perpendicular components of the magnetic field are the only ones affected by the shock compression. The largest increase in both runs is $\sim 38$\% (with respect to the corresponding dashed gray vertical line in Fig.~\ref{fig:std}).
\end{enumerate}
Finally, we show in Fig.~\ref{fig:mag_profile} the profiles of the total magnetic field strength for all runs for two snapshots. These 1D profiles are obtained integrating along the LOS and then taking an average. Each profile is again normalised to the pre-shock magnetic field of each run, $B_{\rm pre}$. The two times shown in Fig.~\ref{fig:mag_profile} correspond to when the shock is starting to compress the turbulent medium (left panel of Fig.~\ref{fig:mag_profile}), and when the shock has crossed the whole simulation box (right panel of Fig.~\ref{fig:mag_profile}). The overall magnetic amplification in the downstream region differs in each run. The discrepancies observed in Fig.~\ref{fig:mag_profile} can be explained by the different distribution of incidence angles between the upstream magnetic field and the shock normal along the shock front in the different runs. The downstream region in the $\mathcal{M}=3$ case develops very similar profiles for both types of turbulence. In particular, the $L/4$ case leads to higher magnetic amplification (compare blue and red lines), whereas the $2L/3$ case leads to comparable downstream magnetic profiles (compare purple and orange lines). We observe that a lower Mach number, such as $\mathcal{M}=2$, leads to less magnetic amplification, owing to the lower shock compression factor. In this case, the final extent of the region where a magnetic field amplification occurs in the downstream is $\sim 25$ kpc larger than in the case with a $\mathcal{M}=3$ shock. This means that the strength of this shock is insufficient to further compress the turbulent medium, thus producing a more extended turbulent magnetic region which will be also reflected in the synchrotron emission.

\subsection{Emission}
\label{sec:emission}
%
\begin{figure}
    \centering
    \includegraphics[width=0.95\columnwidth]{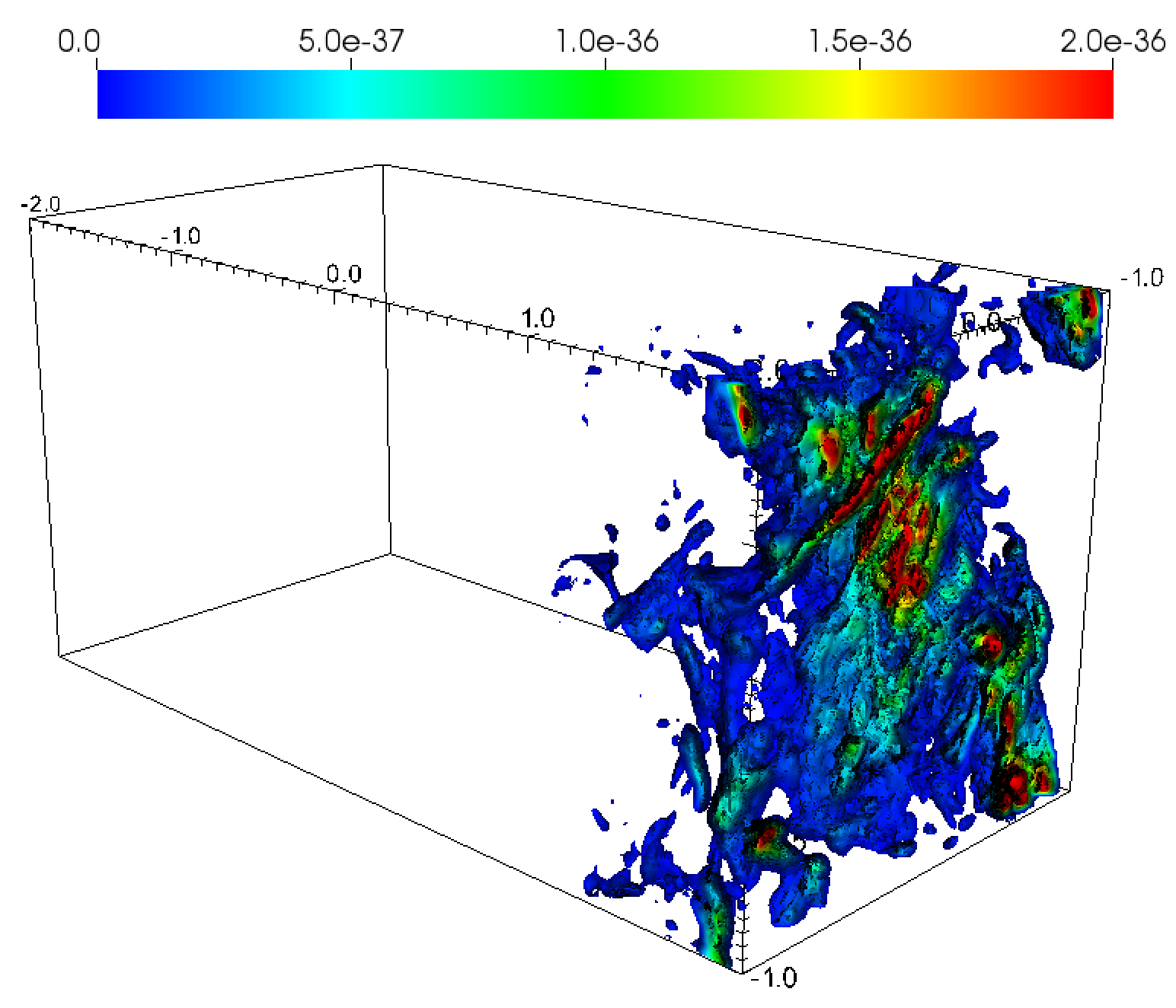}
    \includegraphics[width=0.95\columnwidth]{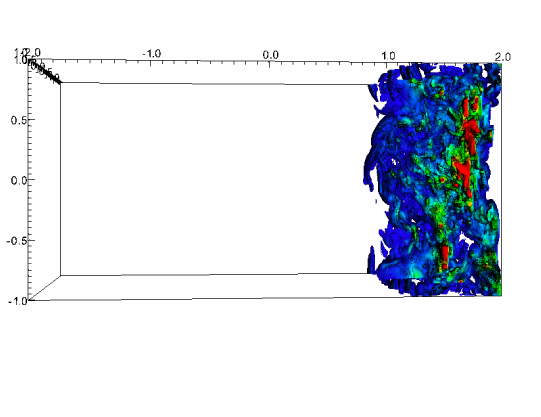}
    \includegraphics[width=0.9\columnwidth]{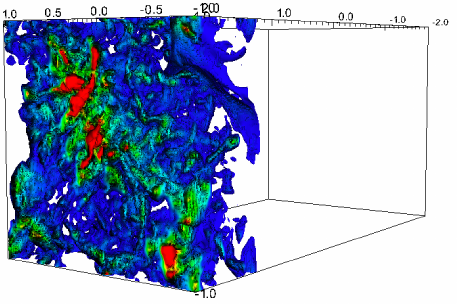}
    \caption{Visualization of synchrotron emissivity $\mathcal{J}_{\rm syn}$ isocurves for the $2L/3$, $\mathcal{M}=3$ and $\theta_{bn}=0^{\circ}$ run at $t=178$ Myr. The emissivity is shown in units of [erg cm$^{-3}$ s$^{-1}$ Hz$^{-1}$ str$^{-1}$]. The axis are shown in units of [100 kpc]. 
    }
    \label{fig:emiss_3d}
\end{figure}

\begin{figure*}
  \centering
  \includegraphics[width=.33\textwidth]{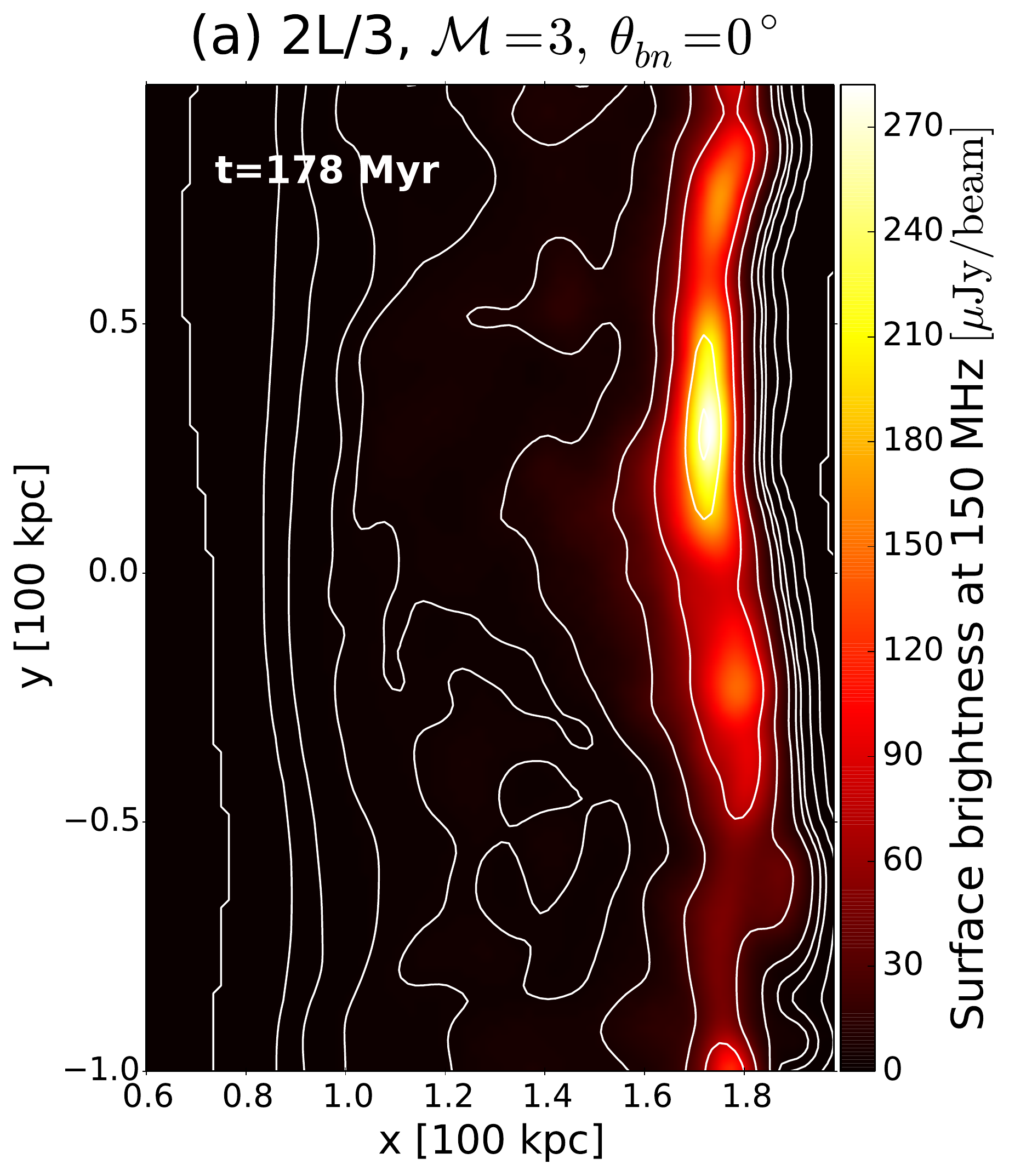}
  \includegraphics[width=.33\textwidth]{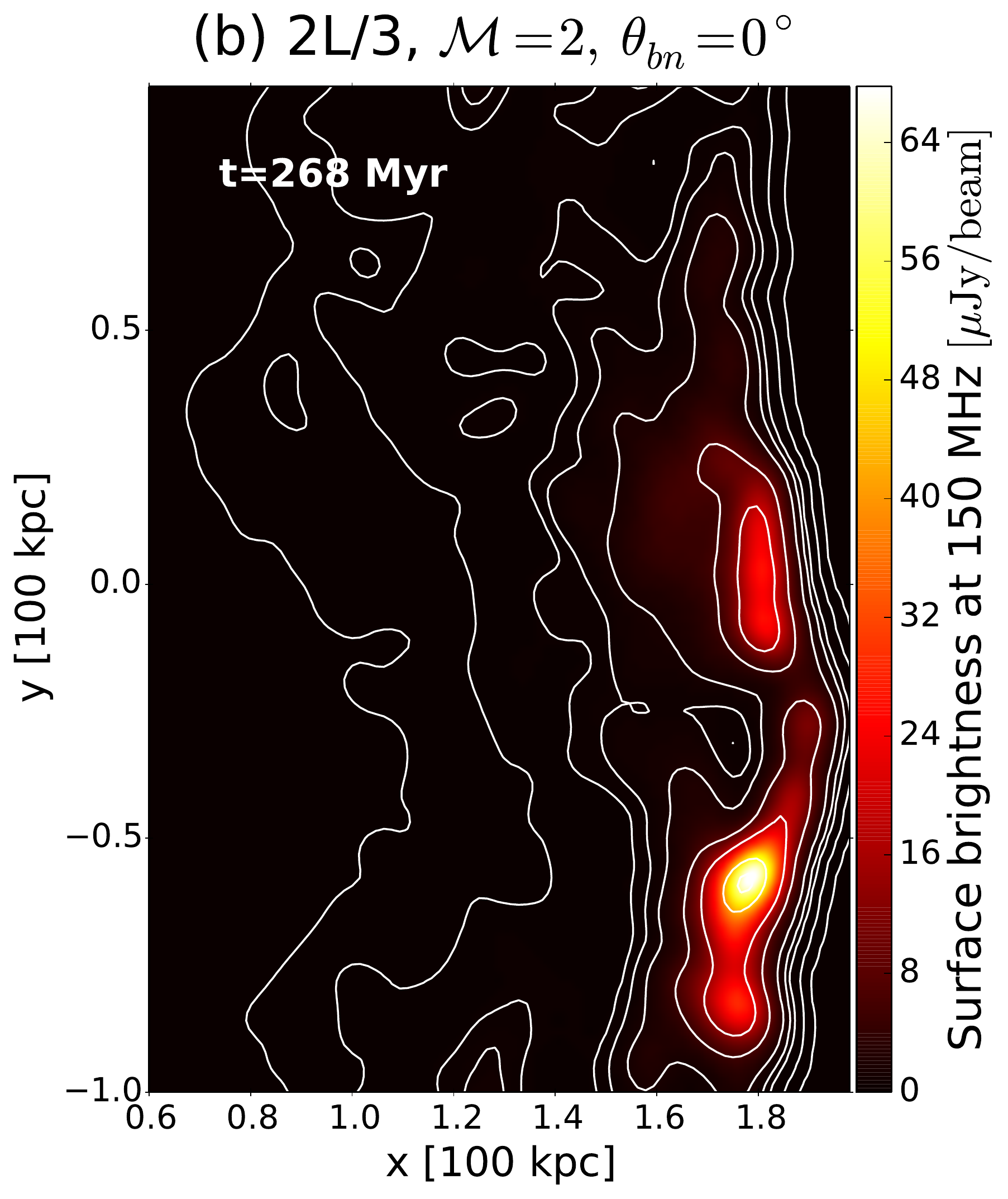}
  \includegraphics[width=.33\textwidth]{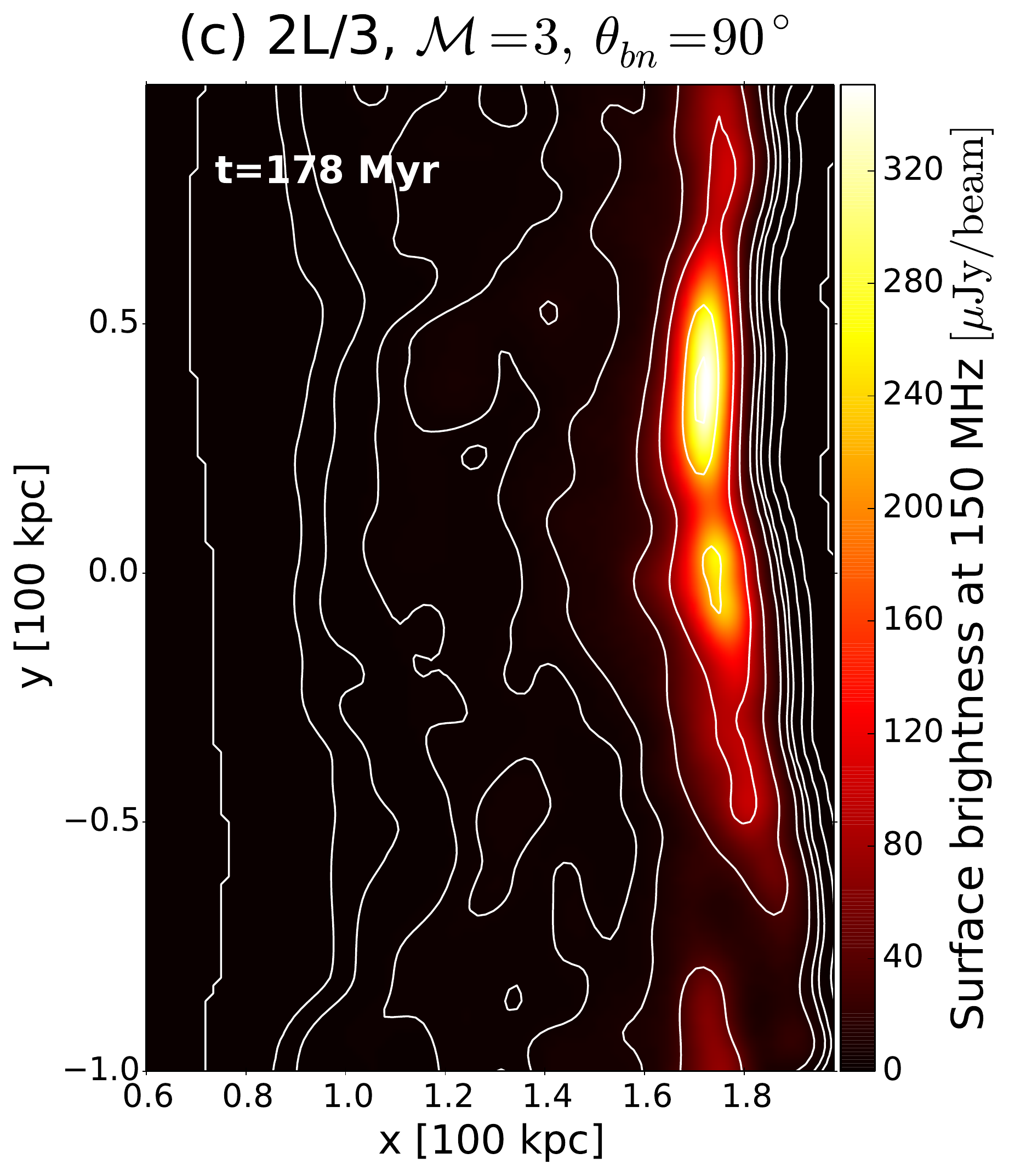} \\
  \includegraphics[width=.33\textwidth]{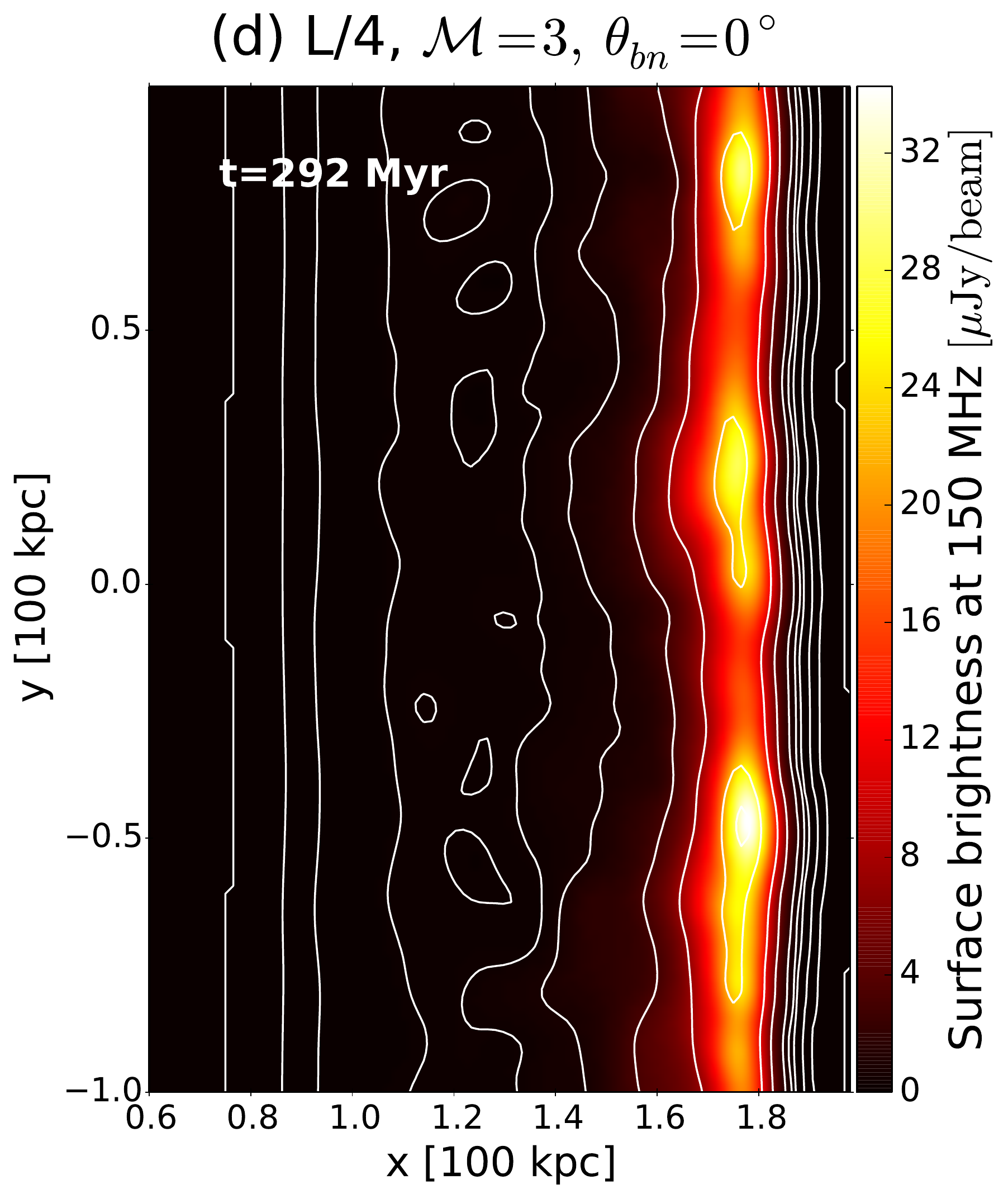}
  \includegraphics[width=.33\textwidth]{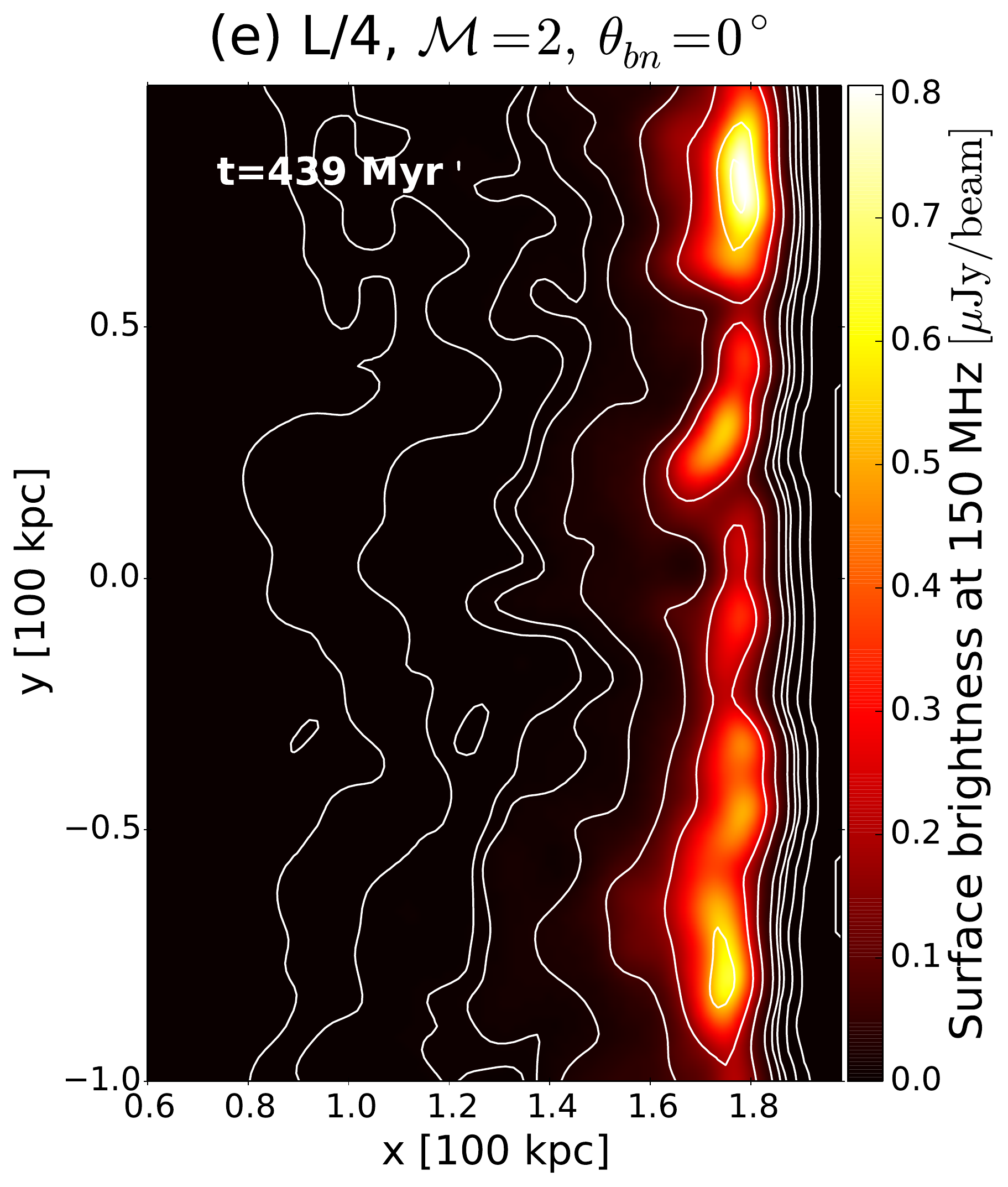}
  \includegraphics[width=.33\textwidth]{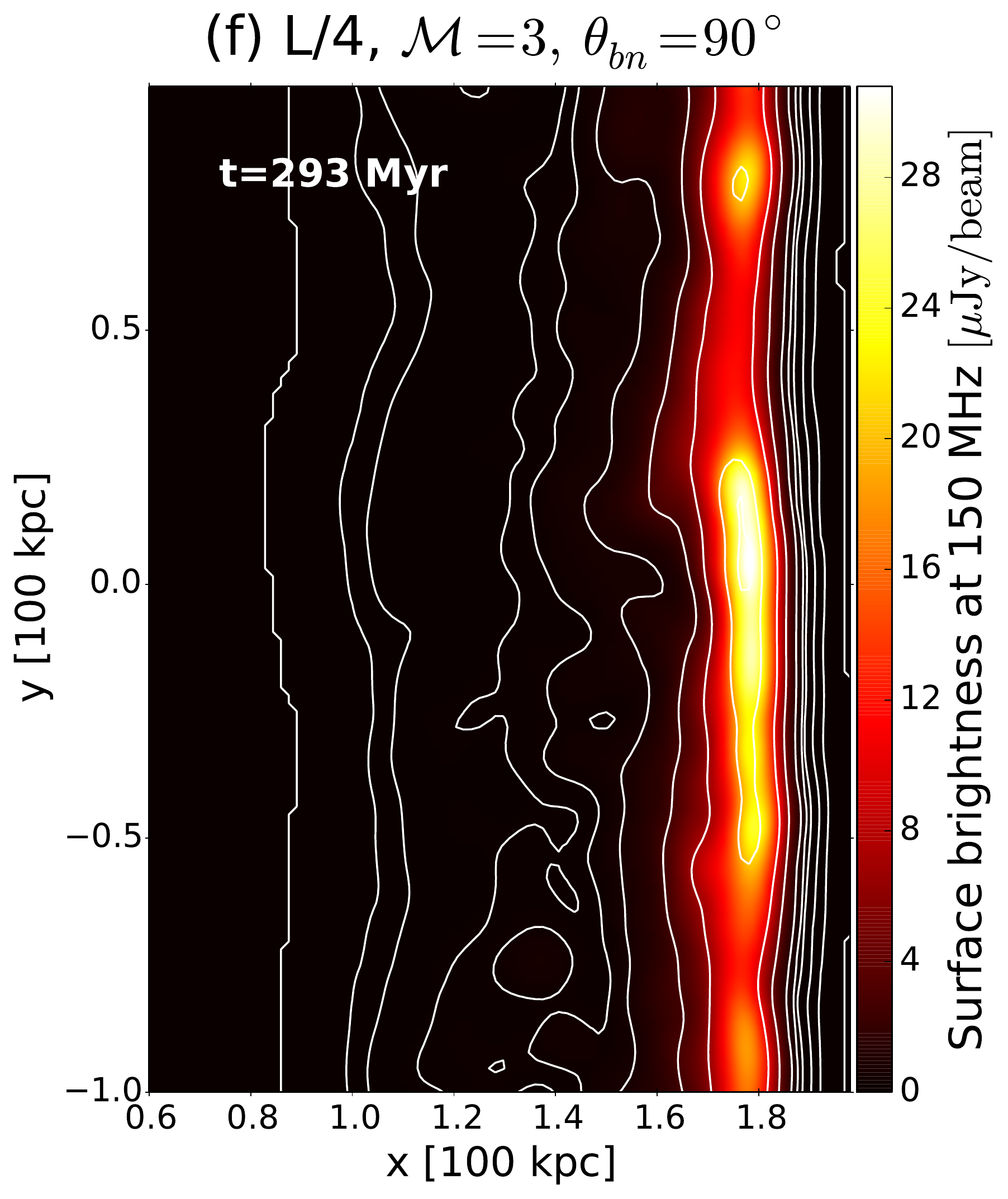}
    \caption{Surface brightness at 150 MHz for all runs in Tab.~\ref{table:init} (see Equation \ref{eq:intensity}). We considered a beam of $\theta^2 = 15"\times15"$ to get the surface brightness ($\theta^2 I_{\nu}$) in units of $\mu$Jy/beam. We smoothed the maps with a Gaussian kernel with $\mathrm{FWHM}=7.24$ kpc (assuming $z=0.023$).}
    \label{fig:emission_maps}
\end{figure*}
We present a few three-dimensional renderings of the synchrotron emission produced by our modelling in Fig.~\ref{fig:emiss_3d}, as seen along different lines of sight. Although the radio emission seems fairly uniform when observed edge-on (see following Fig.~\ref{fig:emission_maps}), the emission is not spatially uniform, but concentrated into threads and filaments in the shock plane. The combination of shock compression and turbulence introduces anisotropies and fluctuations in the flow. This, in turn, directly affects the advection properties and energy evolution of the CR particles. 

We show the surface brightness maps at 150 MHz for all runs at a time when the shock front has reached almost the right end of the simulation box in Fig.~\ref{fig:emission_maps} (the different times are specified in the upper left corner of each panel)\footnote{We have included the surface brightness maps along the x-axis in Appendix~\ref{appen_4} for completeness.}. Note that we only applied Gaussian smoothing for the surface brightness maps in Fig.~\ref{fig:emission_maps} (meant to mimic the finite spatial resolution of a typical LOFAR-HBA observation), while all of our following analysis was done without any smoothing. In the 3D view presented in Fig.~\ref{fig:emiss_3d}, we can distinguish the complex substructure of the emission in the form of filaments, bristles, ribbons or other structures that cannot be classified in a single group. Nevertheless, since the emissivity is not volume-filling and the strength of the emission varies from region to region, some of this structure vanishes in projection. 

The morphology observed in Figs.~\ref{fig:emiss_3d} and \ref{fig:emission_maps} depends on three factors: the strength of the shock, the obliquity of the shock, and the type of turbulence. A higher Mach number leads to stronger emission and elongated patterns due to a stronger compression of the magnetic field. On the other hand, the role of the upstream turbulence is more complex.

One can observe the effect of more elongated patterns for the cases with a shock of $\mathcal{M}=3$ (comparing panels (a), (c), (d) and (f) of Fig.~\ref{fig:emission_maps}) due to the increased stretching of the eddies.

On the other hand, a shock of strength $\mathcal{M}=2$ can produce disrupted patterns in our simulated emission. The shock front is less likely to look totally disrupted in the $L/4$ case (see panel (e)), while this effect is particularly noticeable in the $2L/3$ case (see panel (b)). We notice that this is a direct consequence of the sound speed in the media. For example, the $2L/3$ turbulence easily leads to regions with Mach numbers lower than our threshold ($\mathcal{M}_{\mathrm{min}}=$1.3, see Appendix~\ref{appen_1}).
This happens then because the initial sound speed of the $2L/3$ turbulence is slightly higher than that of the $L/4$ case (see Sec.~\ref{sec:turb}). Subsequently, turbulent dissipation leads to an increase in sound speed and therefore, lowering the Mach number. Such low Mach numbers are not expected to accelerate electrons via the DSA process and therefore, they are excluded from our modelling. In fact, \citealt{Kang_2019} found that quasi-perpendicular shocks with $\mathcal{M}\lesssim 2.3$ may not efficiently accelerate electrons through DSA. Assessing the impact of different turbulent injection scales in $\mathcal{M}=2$ shocks, would then require a tailored set-up which we leave for future work.

Finally, obliquity produces more elongated emission because shock compression only amplifies the component parallel to the shock front (i.e. the $B_y$ and $B_z$ components in this study).

\begin{figure}
    \centering
    \includegraphics[width=0.9\columnwidth]{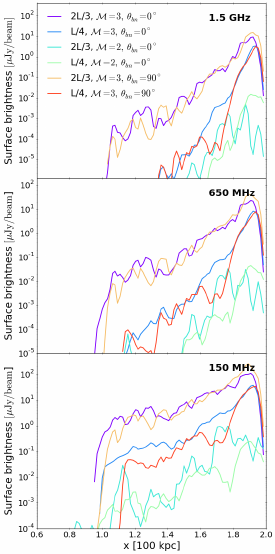}
    \caption{1D surface brightness profiles obtained from the emission maps at 1.5 GHz (\textit{top panel}), 650 MHz (\textit{middle panel}) and 150 MHz (\textit{bottom panel}) for all runs.}
    \label{fig:emission_profiles}
\end{figure}

In Fig.~\ref{fig:emission_profiles}, we show 1D profiles of the emission presented in Fig.~\ref{fig:emission_maps}, at two more observing frequencies: 1.5 GHz and 650 MHz. The most extended emission is found in the runs with $\mathcal{M}=2$ (see bottom panel of Fig.~\ref{fig:emission_profiles}) in agreement with the 1D magnetic profiles previously shown in Fig.~\ref{fig:mag_profile}. These low Mach number runs also show the lowest values of surface brightness along the downstream due to the initial normalisation depending on the Mach number. This suggests that, in our survey of merger shocks, only those with $\mathcal{M}\sim 2$ would require a higher acceleration efficiency to reach observable values as it has been pointed out in previous works \citep[e.g.][]{2020A&A...634A..64B}. Finally, the steepness of the emission profile depends on the magnetic morphology. Comparing all the $\mathcal{M}=3$ runs, the $2L/3$ turbulence case shows a shallower decline compared to the $L/4$ case. This is a direct consequence of the initial distribution of the magnetic field strength. In Fig.~\ref{fig:PDFs_B} of Sec.~\ref{sec:turb}, we showed that initially the $L/4$ run reaches higher magnetic field values in the tail of its PDF, which leads to larger synchrotron losses. 
%
%
\subsubsection{Spectral index}
%
\begin{figure}
    \centering
  \includegraphics[width=0.8\columnwidth]{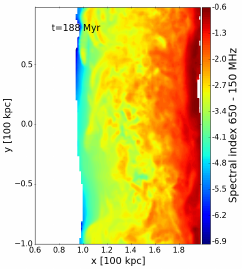} \\
  \includegraphics[width=0.8\columnwidth]{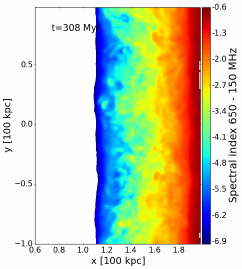}
    \caption{Spectral index maps obtained between 150 MHz and 650 MHz at $t_{\rm shock}$ using Equation (\ref{eq:spectral_index}). \textit{Top panel}: $2L/3$, $\mathcal{M}=3$, $\theta_{bn}=0^{\circ}$ run. \textit{Bottom panel}: $L/4$, $\mathcal{M}=3$, $\theta_{bn}=0^{\circ}$.}
    \label{fig:spec_maps}
\end{figure}
In Fig.~\ref{fig:spec_maps}, we show the spectral index maps for the two different turbulent media with the $\mathcal{M}=3$ shock using Eq. (\ref{eq:spectral_index}). We can observe the expected spectral index gradient starting from the shock front (red) to the end of the downstream region (blue). In the $2L/3$ case, the spectral index values range between $\alpha=-0.6$ and $\alpha=-4.9$, while it goes from $\alpha=-0.6$ to $\alpha=-6.9$ in the $L/4$ case. This agrees with the previous subsection, where we discussed the emission profiles. In the same way, the turbulent medium with smaller initial fluctuations $L/4$ is more likely to produce a steeper gradient because the initial magnetic field distribution has a larger tail (see Fig.~\ref{fig:PDFs_B}).

\begin{figure}
    \centering
  \includegraphics[width=\columnwidth]{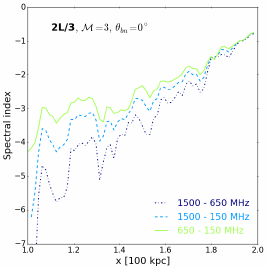}
  \includegraphics[width=\columnwidth]{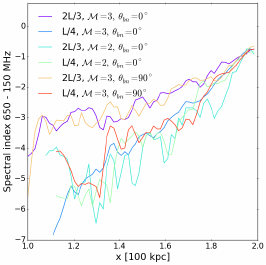}
    \caption{\textit{Top panel}: Spectral index profile for the $2L/3$, $\mathcal{M}=3$, $\theta_{bn}=0^{\circ}$ case at $t=178$ Myr. \textit{Bottom panel}: Spectral index profiles between 150 MHz and 650 MHz for all the cases. The profiles are computed at the same times as in Fig.~\ref{fig:emission_maps}.}
    \label{fig:spec_downstream}
\end{figure}

In Fig.~\ref{fig:spec_downstream}, we show the corresponding spectral index profiles along the downstream region for completeness. In the top panel of Fig.~\ref{fig:spec_downstream} we show how the profile changes when taking into account different frequencies for one specific run: $2L/3$, $\mathcal{M}=3$, $\theta_{bn}=0^{\circ}$. The profiles start to differ beyond $\sim 20$ kpc from the shock front where the emission at lower frequencies decreases more slowly. In the lower panel, we show the spectral index profiles for all of our runs between 650 and 150 MHz. The selected snapshots correspond to those in Fig.~\ref{fig:emission_maps}. For the higher Mach number (i.e. $\mathcal{M}=3$), the $2L/3$ profiles agree more with observations than the $L/4$ profiles. In the $L/4$ case, the spectral index profiles are steeper than observed. For example, the relic in the cluster MACS J0717.5+3745 \citep[see][]{van_Weeren_2017,2018MNRAS.478.2927B} with a Mach number of $\mathcal{M}=2.7$ (inferred from the injection spectral index $\alpha$), shows a spectral index steepening up to values of $\sim -2.5$ over a region of $\lesssim 170$ kpc. Another example is the ``Toothbrush" relic \citep[see][]{2018ApJ...852...65R,2020A&A...636A..30R} which steepens also up to values of $\sim -2.5$ over a region of $\sim 500$ kpc. This suggests that the initial turbulent magnetic field distribution before a shock crossing is rather narrow. For the forced turbulence used in this work, this means that the standard deviation must be smaller than $\sigma_{B}\sim 1 \, \mu$G (see Fig.~\ref{fig:PDFs_B} in Sec.~\ref{sec:turb} for the initial magnetic field distribution). On the other hand, this also suggests that the injection scale (and also the magnetic coherence scale) of the turbulence in galaxy clusters outskirts could be $2L/3$ ($\sim$ 133 kpc) or even larger.

\begin{figure}
    \centering
    \includegraphics[width=\columnwidth]{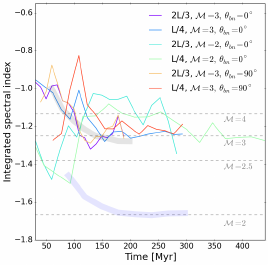}
    \caption{Integrated spectral index evolution computed from fitting the total flux. The gray horizontal dashed lines show the expected integrated spectral index assuming that $\alpha_{\rm int}=\alpha + 1/2$ for different Mach numbers (see Eq. \ref{eq:alpha_int}). The shaded areas correspond to the uniform runs with $\mathcal{M}=2$ (\textit{blue}) and $\mathcal{M}=3$ (\textit{gray}) shocks.}
    \label{fig:spec_fit}
\end{figure}

It is fairly common in observations of radio relics, that only the integrated spectral index can be computed. This is done by fitting the total observed flux (see Eq. \ref{eq:flux}) against different available frequencies. We computed the integrated spectral index in this fashion using 1.5 GHz, 650 MHz and 150 MHz frequencies. In Fig.~\ref{fig:spec_fit} we show how the integrated spectral index evolves as the shock sweeps across the simulation box. In the first $\sim$60 Myr the integrated spectral indices differ quite significantly, whereas after $\sim$140 Myr the value of the integrated spectra seems to converge to the same value for all runs. The relation between the real spectral index $\alpha$ and the integrated one is often assumed to be \citep[e.g.][]{1962SvA.....6..317K,1987MNRAS.225..335H}:
\begin{equation}\label{eq:alpha_int}
    \alpha_{\rm int} = \alpha + \frac{1}{2} = \frac{\mathcal{M}^2 + 1}{\mathcal{M}^2 - 1},
\end{equation}
and therefore we also plot the expected $\alpha_{\rm int}$ for different Mach numbers as a reference in Fig.~\ref{fig:spec_fit} with gray dashed horizontal lines. The integrated spectral index from our runs does not follow a characteristic pattern and it does not remain strictly constant through a time span of roughly 200--300 Myr. In addition, we include the corresponding evolution of the integrated spectral index for two extra runs with a completely uniform medium (density, velocity, pressure and magnetic field) with shocks of strength $\mathcal{M}=2$ and $\mathcal{M}=3$. In the uniform media, Eq. (\ref{eq:alpha_int}) holds after $\sim 50$ Myr when the energy spectrum reaches a steady state at the shock (see \citealt{2017ApJ...840...42K} for a one-dimensional uniform media study), however this is not the case for all the turbulent media. In particular, the effect of turbulence on the 3D distribution of the synchrotron emissivity is that it makes it patchy and not volume-filling (as can be observed in Fig.~\ref{fig:emiss_3d}). This in turn has an effect on the integrated flux and therefore, the integrated spectral index.

Hence, it can be difficult to recover the real spectral index and Mach number through this method, even if recent high-resolution observations of radio relics show consistent integrated and injection spectra \citep[e.g.][]{2020A&A...636A..30R}. The fact that Eq. (\ref{eq:alpha_int}) only holds for a planar shock where all the fields are uniform, suggests that one should be careful when making use of it. One good way to confirm if it is applicable or not is by cross-checking with the result from high-resolution spectral index maps.

This has also been found in previous studies. For example, \citealt{2015JKAS...48....9K} showed that the relation in Eq. (\ref{eq:alpha_int}) only holds for planar shocks, but not for spherical shocks. In the presence of a turbulent medium, the geometry of the shock is more complicated leading to the evolution observed in Fig.~\ref{fig:spec_fit}.

We find that the integrated spectral index in Fig.~\ref{fig:spec_fit} is biased towards higher Mach numbers for all runs. The main reason for this is that the brightest radio emitting regions correspond to the strongest shock compression regions, which in average gives a bias towards higher Mach numbers. In Sec.~\ref{sec:mach}. we will show an analysis of the Mach number distribution and compare it to the one inferred from the thermal fluid. \\

\begin{figure}
    \centering
    \includegraphics[width=\columnwidth]{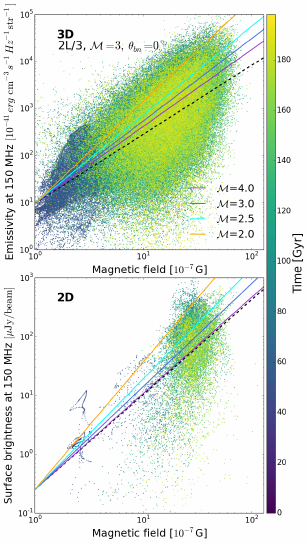}
    \caption{Phase-plots of the magnetic field versus the radio emission at 150 MHz at the cells where the shock front is located for the $2L/3$, $\mathcal{M}=3$, $\theta_{bn}=0^{\circ}$ run. \textit{Top panel}: values extracted out of the 3D distributions, i.e. the emissivity $\mathcal{J}_{\rm syn}$. The coloured lines show the expected fit for different Mach numbers and the black dashed line shows a fit of the data at $t=188$ Myr. \textit{Bottom panel}: values extracted out of the 2D maps, i.e here we have the surface brightness and the values of the magnetic field weighted with the emission (see Eq. \ref{eq:Bweighted}).}
    \label{fig:PP-B}
\end{figure}

In Fig.~\ref{fig:PP-B} we show how the emission at the shock front correlates with the magnetic field strength in 3D and 2D.  We show the  evolution for the $2L/3$ case with $\mathcal{M}=3$ and $\theta_{bn}=0^{\circ}$ run. The emissivity computed from Eq. (\ref{J_syn}) scales with the magnetic field and frequency as 
\begin{equation}
    \mathcal{J}_{\rm syn} \propto B^{(p+1)/2} \nu^{-(p-1)/2},
\end{equation}
\citep[see][]{1979PhB....30..158E} where $\alpha = (p-1)/2$ is the spectral index and $p$ is related to the Mach number through Eqs. (\ref{eq:energy_dist}) and (\ref{spectral_index}),
\begin{equation}\label{eq:fit_alpha}
 \alpha = \frac{(p-1)}{2} = \frac{(q-3)}{2} = \frac{\mathcal{M}^2 + 3}{2(\mathcal{M}^2 -1)}.   
\end{equation}
In the top panel of Fig.~\ref{fig:PP-B}, we show this relation for different Mach numbers with coloured lines and we add an additional black dashed line corresponding to a fit of all the data points. Overall, we find that there is always a systematic mismatch with the initial real Mach number of the shock.  
During the first $\sim 20$ Myr, the 3D distribution shows a sharper relation coinciding with what is expected from Eq. (\ref{eq:fit_alpha}). Nevertheless, shortly after that there will be a spread in the Mach number and magnetic field distribution along the shock front. This spread will keep changing as a consequence of all the turbulent motions in the medium. The black dashed line shows that the relation is biased towards larger Mach numbers. For the 2D case shown in the bottom panel of Fig.~\ref{fig:PP-B}, we considered a magnetic field weighted with the radio emissivity:
\begin{equation}\label{eq:Bweighted}
   B_w = \frac{ \int B \, \mathcal{J}_{\rm syn} dZ}{ \int \mathcal{J}_{\rm syn} dZ}. 
\end{equation}
 We show the same relation pointed out in the upper panel as a reference. The dashed black line in this case corresponds to the fit of all data points in the 2D map. While in this case we have less data points due to the integral along the LOS, it is interesting to see that the bias towards higher Mach numbers is still there. This suggests that the bias is not due to projection effects. We will further discuss the reason behind this bias in Sec.~\ref{sec:mach}.

\subsection{Spectral properties}

\subsubsection{Magnetic field}
\label{sec:mag_spec}

\begin{figure}
    \centering
    \includegraphics[width=\columnwidth]{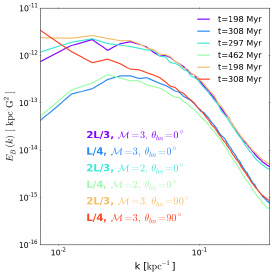}
    \includegraphics[width=\columnwidth]{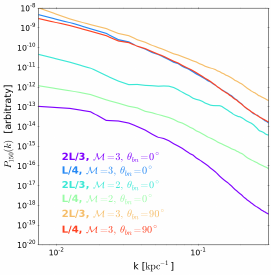}
    \caption{\textit{Top panel}: Final magnetic energy spectrum for all our runs. The spectra are computed in a (200 kpc)$^3$ volume through Eqs. (\ref{eq:energy_spectrum}) and (\ref{eq:power_spectrum}). The final time step differs for each run and it is specified in the legend. \textit{Bottom panel}: Final power spectrum of the synchrotron emissivity at 150 MHz for all our runs (same times as in the top panel).}
    \label{fig:MAGspectrum_all}
\end{figure}

\begin{figure*}
    \centering
    \includegraphics[width=\columnwidth]{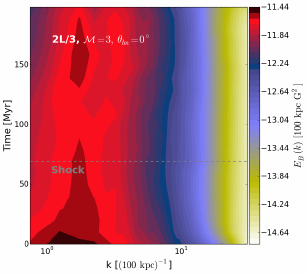}
    \includegraphics[width=\columnwidth]{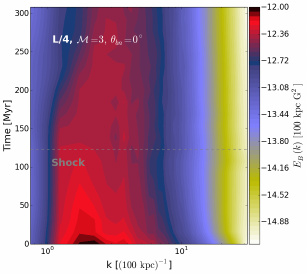}\\
    \includegraphics[width=\columnwidth]{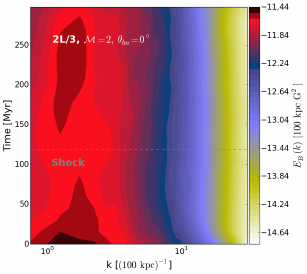}
    \includegraphics[width=\columnwidth]{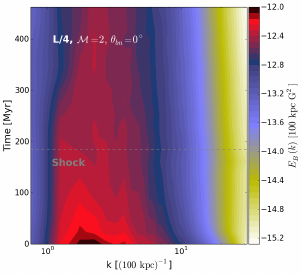}\\
    \includegraphics[width=\columnwidth]{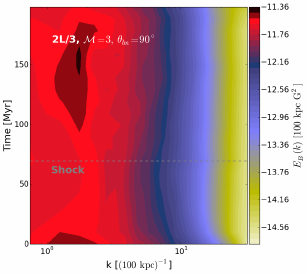}
    \includegraphics[width=\columnwidth]{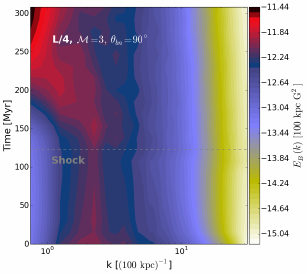}
    \caption{Evolution of the magnetic energy spectrum in the turbulent part of the box (region III [0,2] in Fig.~\ref{fig:init} corresponding to a (200 kpc)$^3$ cube). At each time-step ($y$-axis), we show the magnetic power spectrum by colouring the $x$-axis with its amplitude (see colourbar in logarithmic scale). }
    \label{fig:MAGspectrum_maps}
\end{figure*}

One important feature that characterizes the magnetic field is its \textit{power spectrum}:
\begin{equation}
P_{ij} (\mathbf{k})= \frac{1}{(2\pi)^3}\int \int \int 
e^{-i\mathbf{k} \cdot \mathbf{x}}
R_{ij} (\mathbf{k}) d\mathbf{k},
\end{equation}
where $R_{ij}=\left\langle b_i(\mathbf{x_0})b_j(\mathbf{x_0} + \mathbf{x}) \right\rangle$ 
is the two-point correlation function between the magnetic fields $b_i$ and $b_j$ \citep[e.g.][]{batchelor_1951}. In the case of homogeneous and isotropic fields the relation between the spectral energy and the 1D power spectrum is found to be
\begin{equation}\label{eq:energy_spectrum}
E(k)=2\pi k^2P_{ii}(k).
\end{equation}
We will refer to the 1D power spectrum $P_{ii}(k)$ simply as $P(k)$. In order to obtain the 1D power spectrum we averaged the 3D power spectrum of the magnetic field over spherical shells:
\begin{equation}\label{eq:power_spectrum}
    P(k) = \frac{1}{N_k} \sum_{k-\frac{1}{2}<\mid \mathbf{k} \mid \leq k+ \frac{1}{2}} P(\mathbf{k}).
\end{equation}
We computed the energy power spectrum in the right half part of the simulation (region III, [0,200] kpc) (see Fig.~\ref{fig:init}). In the top panel of Fig.~\ref{fig:MAGspectrum_all} we show the final magnetic energy spectra for all our runs and in the bottom panel we show the final power spectra computed for the emissivity at 150 MHz. In Fig.~\ref{fig:MAGspectrum_maps}, we present the whole evolution of the magnetic energy spectrum $E_B(k)$ condensed in the form maps. Each of these maps contains the following information: the $y$-axis shows the evolution and the $x$-axis shows the wavenumber $k$ coloured with the amplitude of the magnetic energy spectrum at that $k$. In this way, the darker colours denote the regions where the power is peaking (see Fig.~\ref{fig:MAGspectrum_all} as a reference). The dashed gray line is a reference for the reader to know when the shock enters this turbulent region, i.e. region III. We are interested in understanding the effect of shocks in the magnetic energy spectrum, in presence of an ICM with  decaying turbulence. We can see in Fig.~\ref{fig:MAGspectrum_maps} two important features: i) the wavenumbers $k \gtrsim 10$ (corresponding to scales $\lesssim 10$ kpc) are largely unaffected by shocks with strength $\mathcal{M}=$2--3. The resolution may play a role in this case. For example, when a shock enters a turbulent medium, Richtmyer–Meshkov instabilities peaking at small scales can arise. Nevertheless, such instability would take a long time to grow considering weak shocks and therefore, we do not expect those to have a major effect in the context of radio relics;  
ii) the shock compression has a notable effect only in the $\theta_{bn}=90^{\circ}$ cases (see the two bottom panels of Fig.~\ref{fig:MAGspectrum_maps}) at $k\lesssim 2$ (corresponding to scales $\gtrsim 50$ kpc). In these cases, power shifts towards smaller wavenumbers (larger scales) due to the new turbulence introduced after the shock passage. This agrees with previous results from cosmological MHD simulations. In \citealt{2019MNRAS.486..623D}, we observed the same effect when analysing the evolution of a merging galaxy cluster over a time span of almost 10 Gyr. After every merger, shock waves are created and after every shock crossing, the magnetic power shifts to smaller wavenumbers (or larger scales). 

\subsubsection{Emission}
 
In this subsection we make the same analysis as in subsection \ref{sec:mag_spec}, but for the radio emission and we compare it to the results from the magnetic field. We compute the power spectrum of the 3D distributions and also of the 2D maps (computed by integrating along the LOS). The 1D power spectrum is obtained by averaging the 3D spectrum over spherical shells (see Eq. (\ref{eq:power_spectrum})) as mentioned above; while for the 2D maps, we averaged the 2D spectrum over annuli. Afterwards we can compute the characteristic length of the power spectrum for the emission as
\begin{equation}\label{eq:lambda}
    \lambda_c = \frac{\int k^{-1}P(k) \,dk}{\int P(k) \,dk},
\end{equation}
and for the magnetic field as,
\begin{equation}
    \lambda_B = \frac{\int k^{-1}P_B(k) \,dk}{\int P_B(k) \,dk},
\end{equation}
where $P(k)$ and $P_B(k)$ correspond to the power spectrum of the synchrotron emission (see Eq. (\ref{J_syn})) and the power spectrum of the magnetic field, respectively.

\begin{figure*}
    \centering
    \includegraphics[width=\columnwidth,height=12cm]{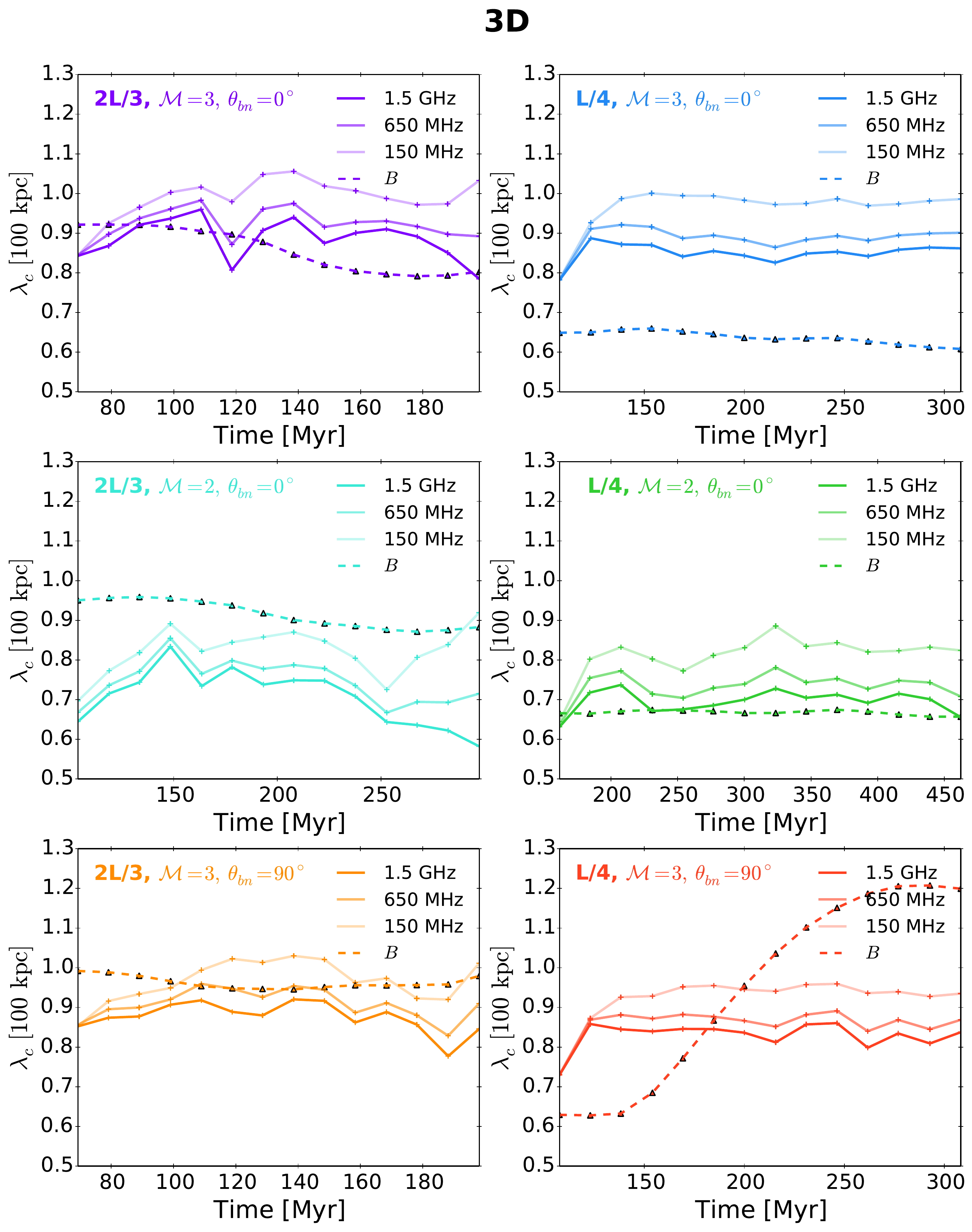}
    \includegraphics[width=\columnwidth,height=12cm]{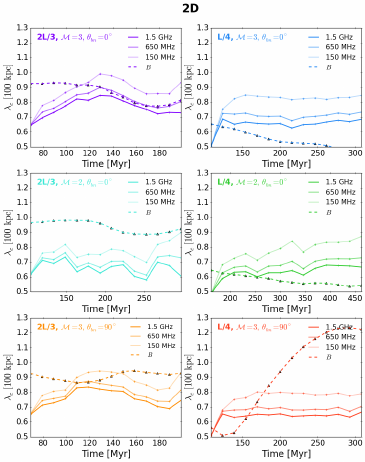}
    \caption{\textit{First two columns}: Characteristic scale of the 3D distributions according to Eq. (\ref{eq:lambda}). \textit{Last two columns}: Characteristic scale of the 2D distributions. Each panel shows the evolution of $\lambda_c$ for 1.5 GHz, 650 MHz and 150 MHz and also the evolution of the magnetic field's characteristic scale $\lambda_B$.}
    \label{fig:lambda_3d}
\end{figure*}

In the first two columns of Fig.~\ref{fig:lambda_3d} we show the results for the 3D case for all of the runs including also the characteristic scale of the magnetic field $\lambda_B$. The characteristic scale of the radio emission is in general of the same order of the characteristic scale of the magnetic field, that is of the order of $\lesssim$100 kpc. There are some specifics regarding 1) the Mach number: higher Mach numbers lead to larger emission scales, for example a $\mathcal{M}=3$ leads to a maximum scale of $\sim 100$ kpc, whereas a $\mathcal{M}=2$ leads to a maximum of $\sim 80$ kpc; and 2) the injection scale of the turbulence in the $\theta_{bn}=0^{\circ}$ case: the characteristic scale of the emission seems not to be affected by the injection scale, but it cannot be directly correlated with the magnetic field's scale. For example, the top panels of the first two columns in Fig.~\ref{fig:lambda_3d} show how $\lambda_c$ is of the same order for both cases while the underlying turbulence is different. In contrast, $\lambda_B$ is directly affected by the underlying turbulence and therefore, it is of a different order. The type of injection scale of the turbulence in the $\theta_{bn}=90^{\circ}$ case plays a big role for the evolution of $\lambda_B$, but this is not reflected in the evolution of $\lambda_c$. This happens only due to the fact that the acceleration efficiency, $\eta$, does not depend on $\theta_{bn}$ in our modelling. The role of a varying $\eta$ will be subject of future work.\\
The characteristic scales of the integrated LOS variables are also shown in Fig.~\ref{fig:lambda_3d} (last two columns). In this case, $\lambda_B$ can be smaller than in reality is by roughly 17--23\% in some cases, while $\lambda_c$ is only smaller by $\sim$10-15 \%. Therefore, we do not observe strong changes in these results due to projection effects.

In summary, we find that the characteristic scales that can be derived from the radio emission could serve as a good proxy for knowing the order of magnitude of the magnetic field's characteristic scale. However, we also find a rather complicated evolution that cannot give us a one-to-one correlation between these two scales and therefore, this exercise alone will not give us much information regarding the type of turbulence existing in the outskirts of ICM. In practice, also the resolution of the radio telescopes plays an important role and will definitely affect these results. 
%

\subsection{Mach number distributions}
\label{sec:mach}

\begin{figure*}
    \centering
    \includegraphics[width=0.9\columnwidth]{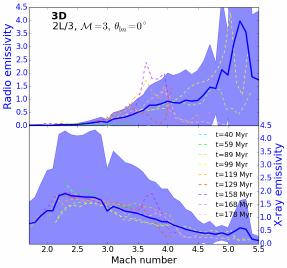}
     \includegraphics[width=0.9\columnwidth]{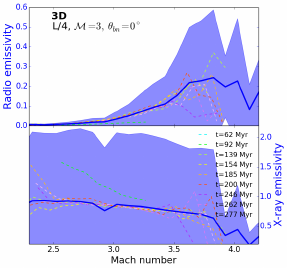} \\
     \includegraphics[width=0.9\columnwidth]{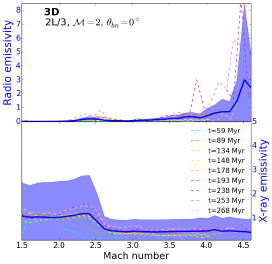}
     \includegraphics[width=0.93\columnwidth]{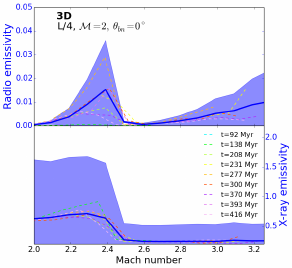} \\
     \includegraphics[width=0.9\columnwidth]{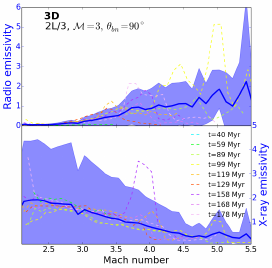}
     \includegraphics[width=0.93\columnwidth]{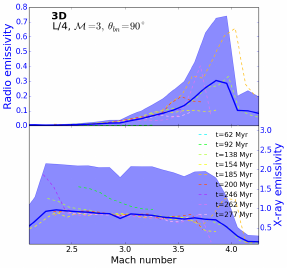} \\
    \caption{Binned statistics of radio emissivity at 1.5 GHz (\textit{left axis}) and X-ray emissivity (\textit{right axis}) with the Mach number of the shock front for different time steps (dashed coloured lines). The dark blue line shows the statistics taken into account the whole  evolution and the shadowed areas denote the corresponding 75-th percentile of the distribution. The emissivity is shown in units of [$10^{-38} \mathrm{erg} \, \mathrm{cm}^{-3} s^{-1} \mathrm{Hz}^{-1} \mathrm{str}^{-1}$].}
    \label{fig:emiss_vs_mach}
\end{figure*}

\subsubsection{3D distribution}

 While the shock front's Mach number distribution peaks at its initial Mach number for almost the whole evolution, it develops a tail towards higher Mach numbers owing to turbulent motions\footnote{see an example for the case $2L/3$, $\mathcal{M}=3$, $\theta_{bn}=0^{\circ}$ in  Fig.~\ref{fig:hist_mach} of the Appendix \ref{appen_1}}. In order to study its impact on the emission, we obtain the emissivity, $\mathcal{J}_{\rm syn}$, corresponding to the cells tracking the shock front at each time step and correlated it with its Mach number. We can do this because the particles are being activated whenever they detect the shock front. In Fig.~\ref{fig:emiss_vs_mach}, we show the binned statistics for the two quantities obtained considering 16 time steps. 

In order to generate an X-ray alike estimate of the Mach number, we computed the X-ray emissivity 
\begin{equation}
    \mathcal{J}_{\rm X-ray} = 1.2 \times 10^{-28} T^{1/2} \left( \frac{\rho}{m_p} \right)^2,
\end{equation}
in units of [erg  cm$^{-3}$ s$^{-1}$ Hz$^{-1}$ str$^{-1}$], where $m_p$ is the proton's mass and $T$ and $\rho$ are in cgs units. Thus, we selected the emission $\mathcal{J}_{\rm syn}$ and $\mathcal{J}_{\rm X-ray}$ only at the shocked cells at each time step in order to compute the binned statistics. We show the distributions at different times with dashed coloured lines and that of the whole evolution with the solid blue lines in Fig.~\ref{fig:emiss_vs_mach}. The top panels show the binned statistics considering the radio emissivity and the bottom panels show that of the X-ray emissivity.

The discrepancy between the two statistical distributions is evident. While the radio emissivity is always biased towards larger Mach numbers, the X-ray emissivity is biased towards smaller Mach numbers. In fact, the peak of the binned distribution of the radio emissivity (temporal envelope) is similar to the values in the tail of the real 3D Mach number distribution (i.e. see an example in Fig.~\ref{fig:hist_mach} in Appendix \ref{appen_1}). Out of the six cases analysed, only the $L/4$, $\mathcal{M}=2$, $\theta_{bn}=0^{\circ}$ case shows a partial match between both distributions. We also observe that the radio-Mach number statistics  fluctuates with time such that the peak Mach number can vary a fair bit. 

In addition, we analysed two runs with uniform media (all fields) but the same Mach numbers (i.e. $\mathcal{M}=2$ and $\mathcal{M}=3$). We verified that in this case the distributions of both emissivities and Mach numbers are the same. Hence, our results suggest that the difference in Mach numbers in radio and in X-ray is a result of turbulence. 

In the presence of turbulence, the radio emissivity is not volume-filling, while the X-ray emissivity comes from the whole shock front. The fact that compression is different from region to region leads to a patchier radio emissivity that can probe a limited part of the shock front (see Fig.~\ref{fig:emiss_3d}). Apart from that, the radio emission is biased towards higher Mach numbers because the initial CR energy is $\propto \mathcal{M}^3$ (which in our work corresponds to the normalisation $N_0$ defined in Eq.(\ref{eq:norm})). Finally, the magnetic field fluctuations created after the shock-crossing play an important role. As discussed in Sec.~\ref{sec:fluid}, the amplitude of the magnetic field fluctuations (which directly affects the radio emission) decreases more
slowly than the velocity, density and temperature fluctuations. This adds to the discrepancy because the X-ray emissivity depends only on the temperature and density fields.

\section{Summary and conclusions}
\label{sec:conclusions}

We have presented a hybrid framework to compute the synchrotron emission from a shock wave propagating through a medium with decaying turbulence representing a small fraction of the ICM. In our framework, the MHD grid represents a thermal fluid, whereas Lagrangian particles represent CR electrons. 
 We injected CR electrons at the shock discontinuity assuming diffusive shock acceleration. Each CR electron evolves according to the cosmic-ray transport equation in the diffusion approximation.

Our simulations explored shocks with Mach numbers characteristic of radio relics, i.e. $\mathcal{M}=2$ and $\mathcal{M}=3$. Moreover, we varied the downstream turbulence using turbulence-in-a-box simulations: a solenoidal subsonic turbulence with power peaking at 2/3 of the box (case $2L/3$) and a solenoidal subsonic turbulence with power peaking at 1/4 of the box (case $L/4$). One snapshot of each simulation was selected as an initial condition for our shock-tube simulation. Our results can be summarized as follows:
\begin{itemize}
    \item[i)] \textit{Impact of a shock on decaying turbulence}: We find that mild shocks produce magnetic fluctuations in the downstream region that do not correlate with fluctuations in velocity, density and temperature. In fact, we find that magnetic fluctuations can increase even when velocity fluctuations decrease. This, in turn, can affect the final extent of the magnetic downstream. We find the strongest effect in perpendicular shocks, as expected from theory. Shocks with $\mathcal{M}=2$ travelling in a medium with smaller fluctuations, such as our $L/4$ case, cause the least effect and seem to hardly modify the initial magnetic field distribution. \\
    
    \item[ii)] \textit{Radio emission}: The existence of substructure in the synchrotron emission is a direct consequence of a turbulent medium. We found that $\mathcal{M}=2$ shocks in our set-up are unlikely to reproduce observable radio relics. 
    The physical reason behind this is that $\mathcal{M}=2$ shocks are not strong enough to modify the initial pre-shock magnetic field. For example, the relic at Abell 2744 \citep[e.g.][]{Govoni_2001,Eckert:2016ubk,2017ApJ...845...81P,Paul_2020} reaches a surface brightness of the order of tens of mJy/beam at 1.4 GHz, which would require an acceleration efficiency of $\eta \sim 1$ in our $L/4$ turbulence. \\

    \item[iii)] \textit{Discrepancies in the spectral index}: Our spectral index profiles suggest that in the case of $\mathcal{M}=3$ shocks, a turbulent injection scale of $2L/3$ or even larger reproduces observations better than the $L/4$ case. The $L/4$ initial magnetic field distribution allows for higher values of the magnetic field strength reflected in the tail of the PDF which steepens the spectral index profiles more than in the $2L/3$ case. We conclude that an initial turbulent, magnetic field distribution in the ICM must have a standard deviation smaller than $\sigma_B = 1 \, \mu$G. We compare our results of the integrated spectral index to the relation $\alpha_{\rm int}=\alpha +1/2$ and find that this relation does not seem to hold in the presence of a turbulent medium. The reason for this is the distribution of Mach numbers within a shock front in a turbulent medium. As a consequence, the injected electrons will have different initial energy spectra.\\

    \item[iv)] \textit{Discrepancies in Mach numbers}: We find that the synchrotron emission is biased towards larger Mach numbers when comparing to the X-ray emission. This agrees with previous numerical work \citep[e.g.][]{2015ApJ...812...49H} and a number of observations of radio relics. For example, X-ray observations of the Toothbrush relic in the cluster 1RXS J0603.3+4214 infer a Mach number of $\mathcal{M}\sim 1.5$ \citep[e.g.][]{Ogrean_2013,van_Weeren_2016}, while radio observations infer a higher Mach number of $\mathcal{M}\sim 3.7$ \citep[e.g.][]{2018ApJ...852...65R,2020A&A...636A..30R}. The source of this discrepancy lies in 1) the stronger dependence of the synchrotron emission on the compression in the shock and 2) the fact that the amplitude of the magnetic field fluctuations (which affect the radio emission) decreases more slowly than the density and temperature fluctuations. Hence, higher Mach numbers in the tail of the Mach number distribution bias the overall Mach number.\\
    
    \item[v)] \textit{Magnetic energy spectrum}: We find that scales $\lesssim 10$ kpc are largely unaffected by shocks with $\mathcal{M}=$ 2--3, independent of the type of turbulence. We find that the power shifts towards smaller wave numbers (larger scales) after shock passage which is more pronounced in perpendicular shocks. \\
    
    \item[vi)] \textit{Characteristic length of the radio emission}: The characteristic lengths derived from the power spectrum of the emission, $\lambda_c$, and magnetic field, $\lambda_B$, are of the same order. We find that $\lambda_c$ is in general of the order of $\lesssim 100$ kpc. Analysing the LOS variables, we do not observe strong projection effects and $\lambda_B$ and $\lambda_c$ are only 17--23\% and 10--15\% smaller than in 3D.
\end{itemize}

In summary, we could identify the most important
features that link the observable properties of radio relics with the dynamical properties of the upstream ICM.
Our work confirms that the Mach numbers inferred from the radio emission are likely to be overestimates of the real Mach number of the thermal fluid in the presence of turbulence. This has been previously pointed out as a possible solution that can alleviate the problem of acceleration efficiencies and as a possible explanation for the non-detection of $\gamma$-rays from galaxy clusters (see \citealt{2010ApJ...717L..71A,2014ApJ...787...18A,2016ApJ...819..149A} and \citealt{2014MNRAS.437.2291V}). While CRe and CRp are expected to be accelerated at the same places, their acceleration mechanisms and efficiencies will differ \citep[e.g.][]{2014ApJ...783...91C}. CRe are accelerated preferentially at quasi-perpendicular shocks and CRp at quasi-parallel shocks. Recent works by \citealt{2020MNRAS.495L.112W} and \citealt{2020MNRAS.496.3648B} used cosmological MHD simulations to show that indeed the predominance of quasi-perpendicular shocks in merger and accretion shocks might be enough to explain the absence of CRp. Therefore, in future work we will include the obliquity dependence in our acceleration efficiency, $\eta$. Moreover, we intend to survey a larger range of possible parameters, both, for the ICM conditions and for the shock properties. This will help us to assess whether the large variety of  relic sources can be explained by the model adopted here. We will also present a detailed study considering polarisation (Q and U Stokes parameters, as well as Rotation Measure), different lines-of-sight, projection and beam effects.

\section{Acknowledgements}
 The analysis presented in this work made use of computational resources on the JUWELS
cluster at the Juelich Supercomputing Centre (JSC), under
project "stressicm" with F.V. as principal investigator and P.D.F as co-principal investigator. The 3D visualization of the synchrotron emissivity was done with the VisIt software \citep[see][]{HPV:VisIt}.\\
P.D.F, F.V. and M.B. acknowledge the financial support from the European Union's Horizon 2020 program under the ERC Starting Grant "MAGCOW", no. 714196. W.B.B. acknowledges the financial support from the Deutsche Forschungsgemeinschaft (DFG) via grant BR2026125.\\
We acknowledge our anonymous reviewer for helpful comments on the first version of this manuscript. Finally, we thank D. Ryu, C. Federrath and R. Mohapatra for fruitful scientific discussions.

\section{Data Availability Statement}

The data underlying this article will be shared on reasonable request to the corresponding author.

\bibliographystyle{mnras}
\bibliography{paola,franco}

\appendix

\section{Numerical divergence of the magnetic field}
\label{appen_2}

\begin{figure}
	\includegraphics[width=\columnwidth]{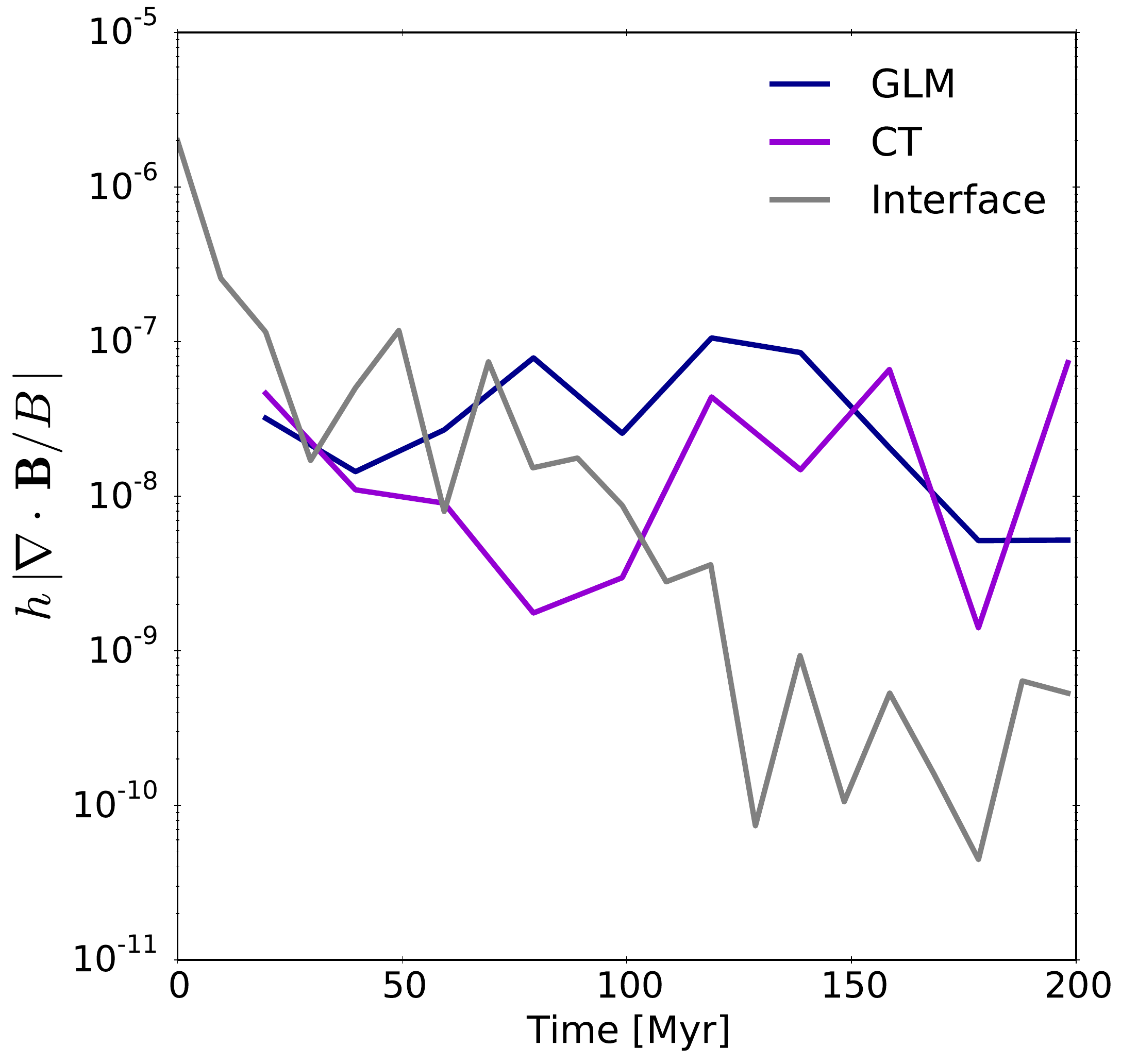}
	\caption{Test on the numerical conservation of the $\nabla \cdot \mathbf{B} = 0$ condition (where $h=\Delta x$). We compare two runs only differing in their divergence cleaning method: GLM (\textit{blue}) and CT (\textit{purple}). The gray line shows the corresponding evolution for the interface between regions II and III using the GLM method. We performed this test using the set-up $2L/3$, $\mathcal{M}=3$ and $\theta_{bn}=90^{\circ}$.}
	\label{fig:divergence}
\end{figure}

The $\nabla \cdot \mathbf{B} = 0$ condition in the particle module of the PLUTO code is maintained with the hyperbolic divergence cleaning technique where the induction equation is coupled to a generalized Lagrange multiplier (GLM) \citep[e.g.][]{2002JCoPh.175..645D}. In \citealt{kri11}, it has been argued that the best results for the divergence-free evolution of the magnetic field are achieved using the constrained transport (CT) method. Keeping this in mind, we tested the differences of the $\nabla \cdot \mathbf{B} = 0$ condition with the GLM and CT techniques. In Fig.~\ref{fig:divergence}, we show the evolution of the magnetic field divergence using the same set-up as our $2L/3$, $\mathcal{M}=3$ and $\theta_{bn}=90^{\circ}$ run (see Table \ref{table:init}) for the whole simulation box. Both methods are comparable and keep the numerical magnetic field divergence under $\sim 0.001$\% of the local magnetic field value. This demonstrates that the use of the GLM divergence cleaning technique is robust against CT for our particular set-up. Thus, we do not expect our numerical scheme to have an impact in our final synchrotron emission maps. Additionally, we show how the divergence condition behaves for the interface between regions II and III in our setup in Fig.~\ref{fig:divergence}. The effects of the initial interpolation of the external input get diminished after a few steps and the numerical magnetic field divergence drops again below $\sim 0.001$\% before the shock enters the region of interest (i.e. region III); see an additional discussion about this in Appendix D of \citealt{2018MNRAS.473.3454B}. 

\section{Shock finder}
\label{appen_1}

\begin{figure}
	\includegraphics[width=\columnwidth]{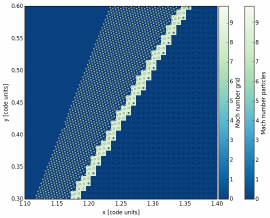}
	\caption{Numerical test on the shock's directionality. We show a 2D parallel shock propagating with an angle of \textbf{$60^{\circ}$}.}
	\label{fig:Mach}
\end{figure}
\begin{figure}
    \centering
    \includegraphics[width=\columnwidth]{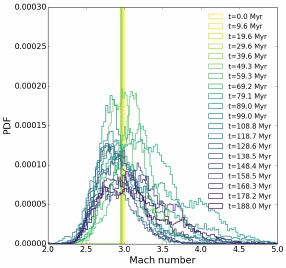}
    \caption{Evolution of Mach number distribution at the shock front for the case $2L/3$, $\mathcal{M}=3$, $\theta_{bn}=0^{\circ}$.}
    \label{fig:hist_mach}
\end{figure}

The algorithm to find shocks is already implemented in the PLUTO code \citep[see][]{2018ApJ...865..144V}. In the first step, shock cells are tagged through the $\nabla \cdot \mathbf{v} < 0$ condition. Then we implemented an extra condition for the activation time of the tracer particles in order to compute the initial Mach number and compression ratio. This in turn is needed for assigning the initial energy spectra with an spectral index $q$ (see Eq. (\ref{spectral_index})) to each tracer particle. The Mach number at the shock center is computed from the Rankine-Hugoniot pressure jump condition:
\begin{equation}
    \Delta \log p \geq \log \frac{p_2}{p_1} \bigg\rvert_{\mathcal{M}=\mathcal{M}_{\rm min}},
\end{equation}{}
where the subscripts 1 and 2 denote the pre-shock and post-shock regions, respectively. The minimum Mach number is set to $\mathcal{M}_{\rm min}=1.3$ and it acts as a threshold to filter out weaker shocks. The pressure jump is computed with the neighbouring cells along the three directions for which a Mach number distribution is finally obtained: 
$\mathcal{M}^2=\mathcal{M}^2_x + \mathcal{M}^2_y + \mathcal{M}^2_z$ (see \citealt{ry03}, \citealt{va11comparison}, \citealt{2015MNRAS.446.3992S}). \\
The PLUTO code is able to compute the compression ratio for the update of the spectra once the particle has crossed a shock, nevertheless we implemented this extra condition as it was necessary for computing the compression ratio at the time of activation and not a time step afterwards as in \citealt{2018ApJ...865..144V}. In this fashion, the tracer particles have an initial spectrum consistent with DSA theory at the time of activation and after a time step their spectra will evolve subject to radiative losses. \\
We tested the directionality of the shock finder by setting up a 2D parallel shock propagating with an angle of $60^{\circ}$ with respect to the $x$-axis and placing one particle per cell on a quarter of the domain. The shock propagates with a Mach number of $\mathcal{M}=10$ and the interpolation of the grid quantity at the particle position in PLUTO is done with standard techniques used for Particle-In-Cell (PIC) codes \citep[see][]{2004CoPhC.164..189B}. We used the Nearest Grid Point (NGP) method for the implementation of this test. In Fig.~\ref{fig:Mach}, we show the Mach grid distribution as well as the particle's interpolated Mach number for a snapshot of this set-up. We also show the evolution of the shock's Mach number in our final set-up for one of our studied cases in Fig.~\ref{fig:hist_mach}.
%

\section{Resolution}
\label{appen_3}

\begin{figure}
	\includegraphics[width=\columnwidth]{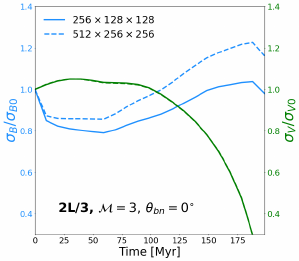}
	\caption{Evolution of the standard deviation of the magnetic field (\textit{left green axis}) and velocity field (\textit{right blue axis}). The solid lines correspond to the lower resolution run ($256 \times 128 \times 128$ cells) and the dashed lines correspond to the higher resolution run ($512 \times 256 \times 256$ cells).}
	\label{fig:resolution1}
\end{figure}
\begin{figure*}
	\includegraphics[width=0.8\columnwidth]{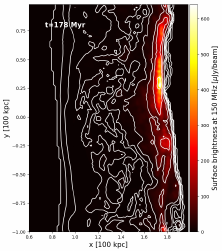}
	\includegraphics[width=0.8\columnwidth]{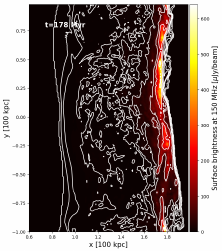} 
	\caption{Surface brightness at 150 MHz for the run $2L/3$, $\mathcal{M}=3$ and $\theta_{bn}=90^{\circ}$. We considered a beam of $\theta^2 = 15"\times15"$ to get the surface brightness ($\theta^2 I_{\nu}$) in units of $\mu$Jy/beam. No smoothing is considered here. \textit{Left panel}: $256 \times 128 \times 128$ cells simulation. \textit{Right panel}: $512 \times 256 \times 256$ cells simulation.}
	\label{fig:resolution2}
\end{figure*}

We tested our set-up doubling the resolution to $512 \times 256 \times 256$ cells and number of Lagrangian particles to 25,165,824. Here we will limit ourselves to show the run $2L/3$, $\mathcal{M}=3$ and $\theta_{bn}=90^{\circ}$. In Fig.~\ref{fig:resolution1} we show the evolution of the standard deviation of the magnetic and velocity field for the resolution used in this work, i.e. $256 \times 128 \times 128$ cells, and for the higher resolution. 
This result was shown and explained in detail in Sec.~\ref{sec:fluid}. This figure confirms that the velocity field (and also the density and temperature field) dynamics is largely unaffected by a higher resolution. In fact, it is only the magnetic field fluctuations that are $\sim 20$\% enhanced. This means that our result on the non-correlation between magnetic and velocity fluctuations is even more accentuated at higher resolution. This is expected as we are increasing the effective Reynolds number of the simulation. An upper limit of the Reynolds number is given by:
\begin{equation}
    {\rm Re}_{\rm max} \approx \left( \frac{l}{ \Delta x} \right)^{4/3},
\end{equation}
where $l$ is the maximum correlation scale in the flow and $\Delta x$ is the resolution. In this case, $l=2L/3\approx 133.3$ kpc, $\Delta x_{256} = 1.56$ kpc and $\Delta x_{512} = 0.78$ kpc would lead to an upper limit of the effective Reynolds number of $\sim 377$ and $\sim 950$ for the low and high resolution runs, respectively. A lower limit is given in contrast by:
\begin{equation}
    {\rm Re}_{\rm min} \approx \left( \frac{l}{\varepsilon \Delta x} \right)^{4/3},
\end{equation}
where $\varepsilon$ is a factor depending on the diffusivity of the numerical method. For second order finite difference/volume codes such as our case with the PLUTO code, one can assume $\varepsilon \approx 7$ \citep[e.g.][]{2011ApJ...737...13K}. This leads to a lower limit of the effective Reynolds number of $\sim 28$ and $\sim 71$ for the low and high resolution runs, respectively.

Finally, we show in Fig.~\ref{fig:resolution2} a comparison between surface brightness maps at 150 MHz with both resolutions. As it can be observed, the broader features as well as the extent of the downstream are consistent in both resolutions, while the higher resolution run highlights smaller features ($\sim 1$ kpc). 
%

\section{Surface brightness along the x-axis}
\label{appen_4}

\begin{figure*}
  \centering
  \includegraphics[width=.33\textwidth]{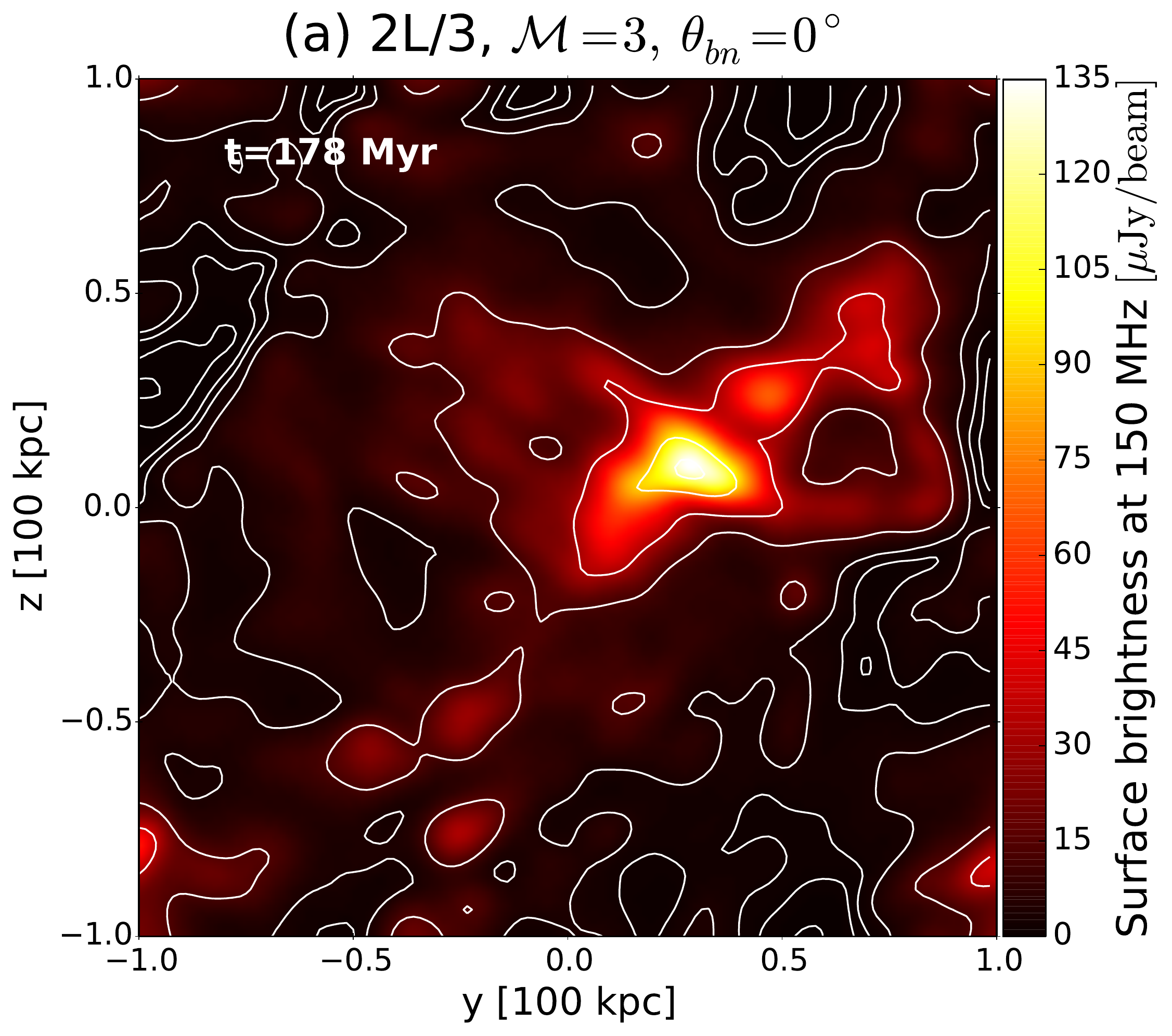}
  \includegraphics[width=.33\textwidth]{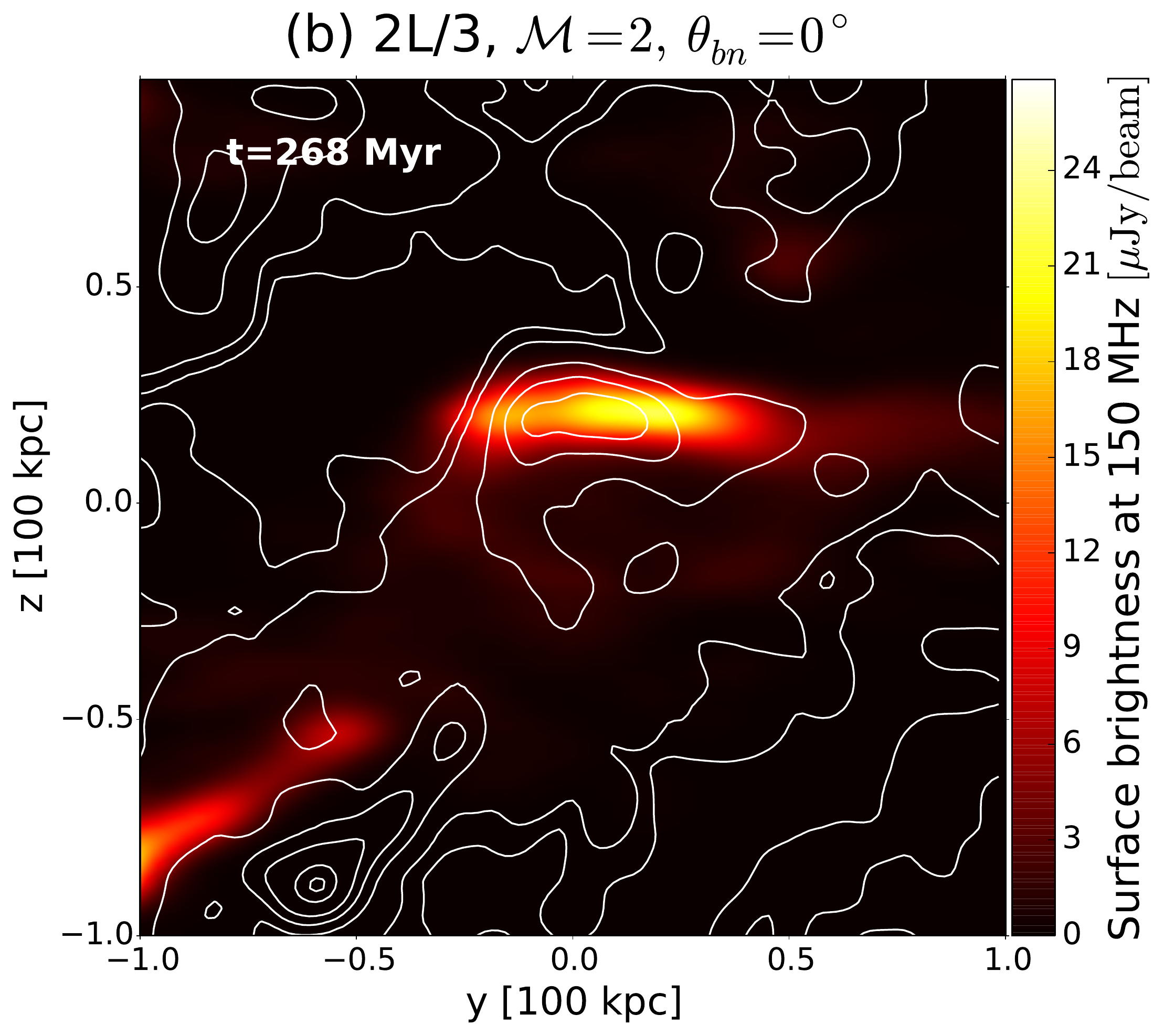}
  \includegraphics[width=.33\textwidth]{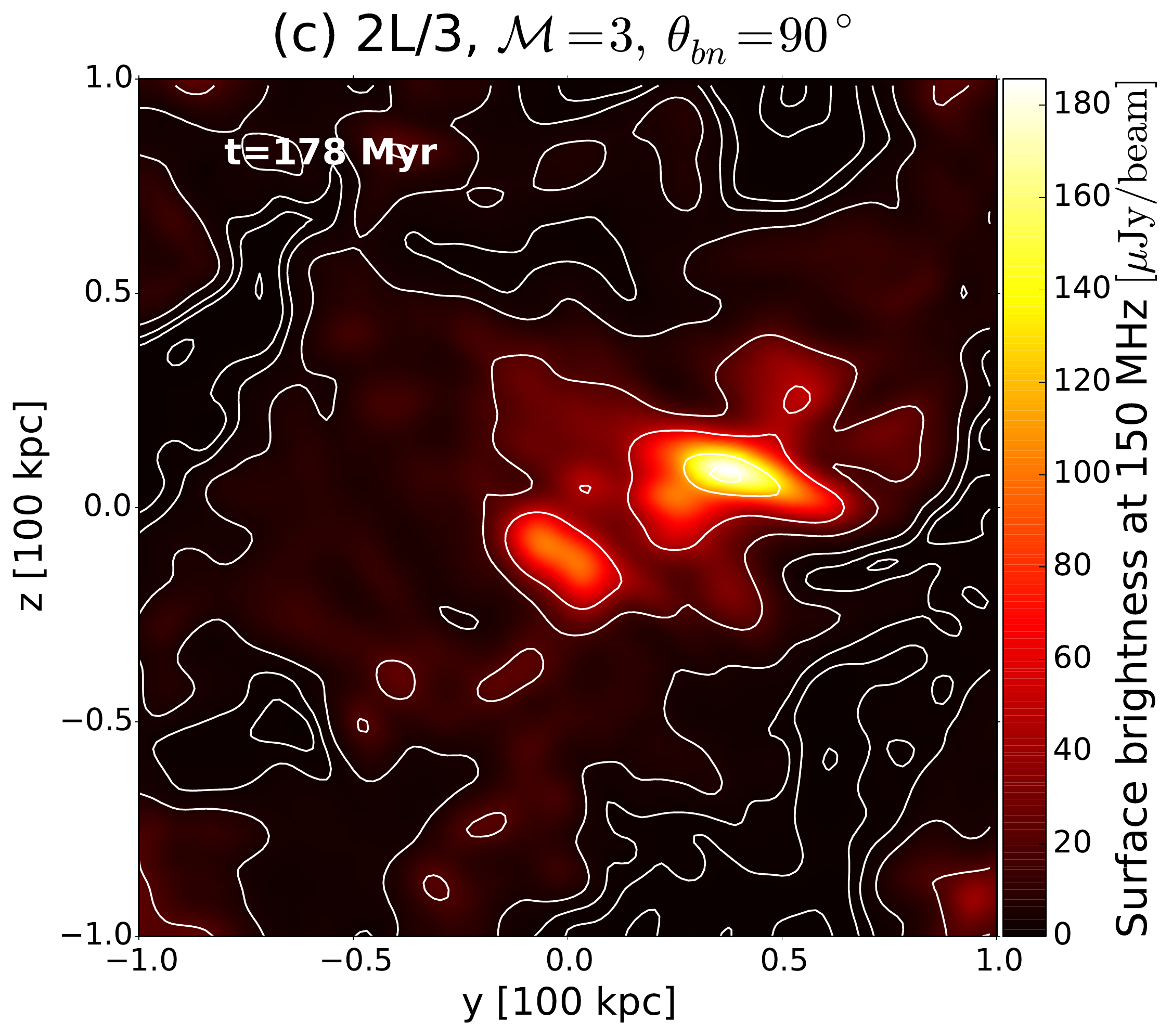} \\
  \includegraphics[width=.33\textwidth]{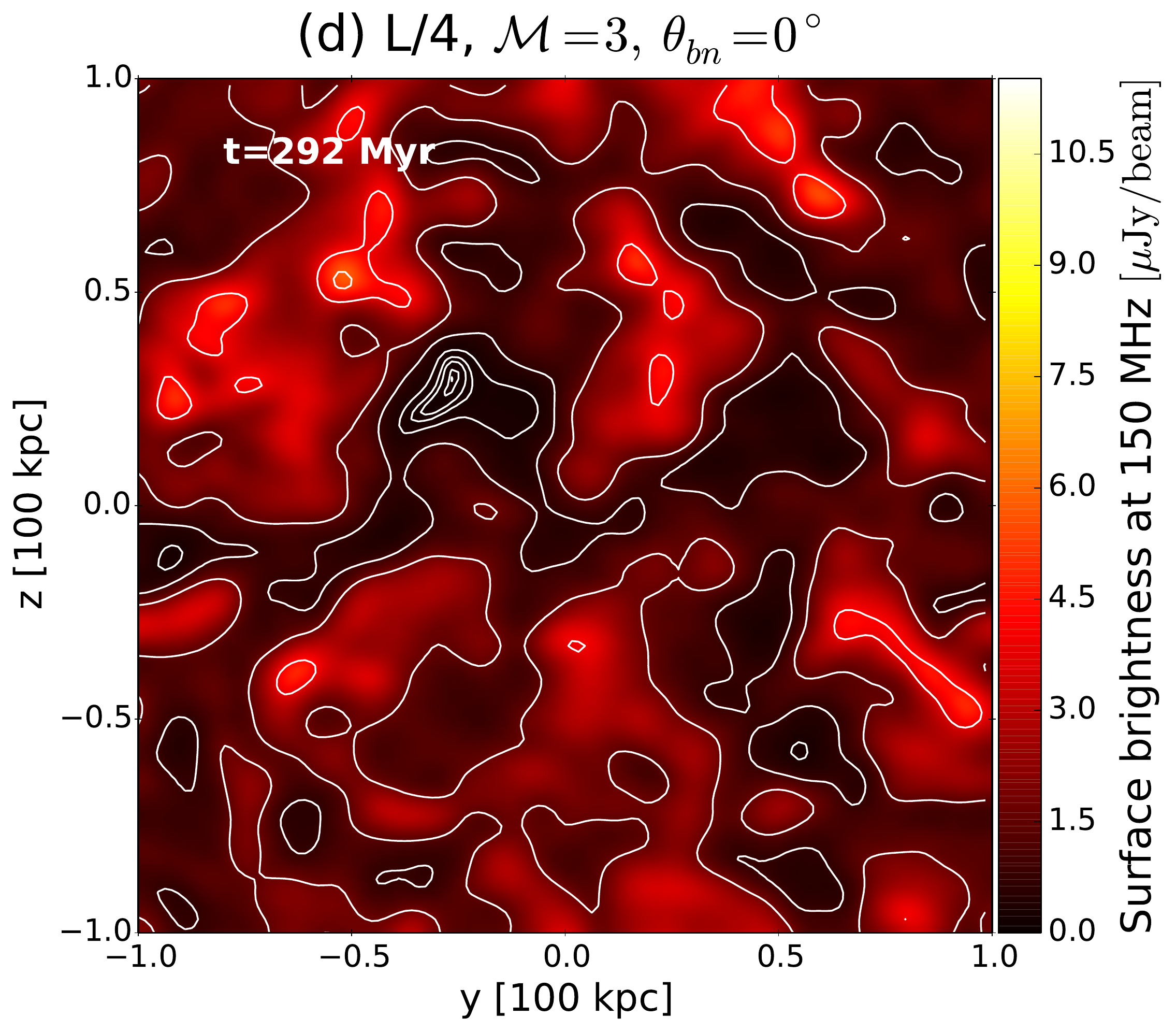}
  \includegraphics[width=.33\textwidth]{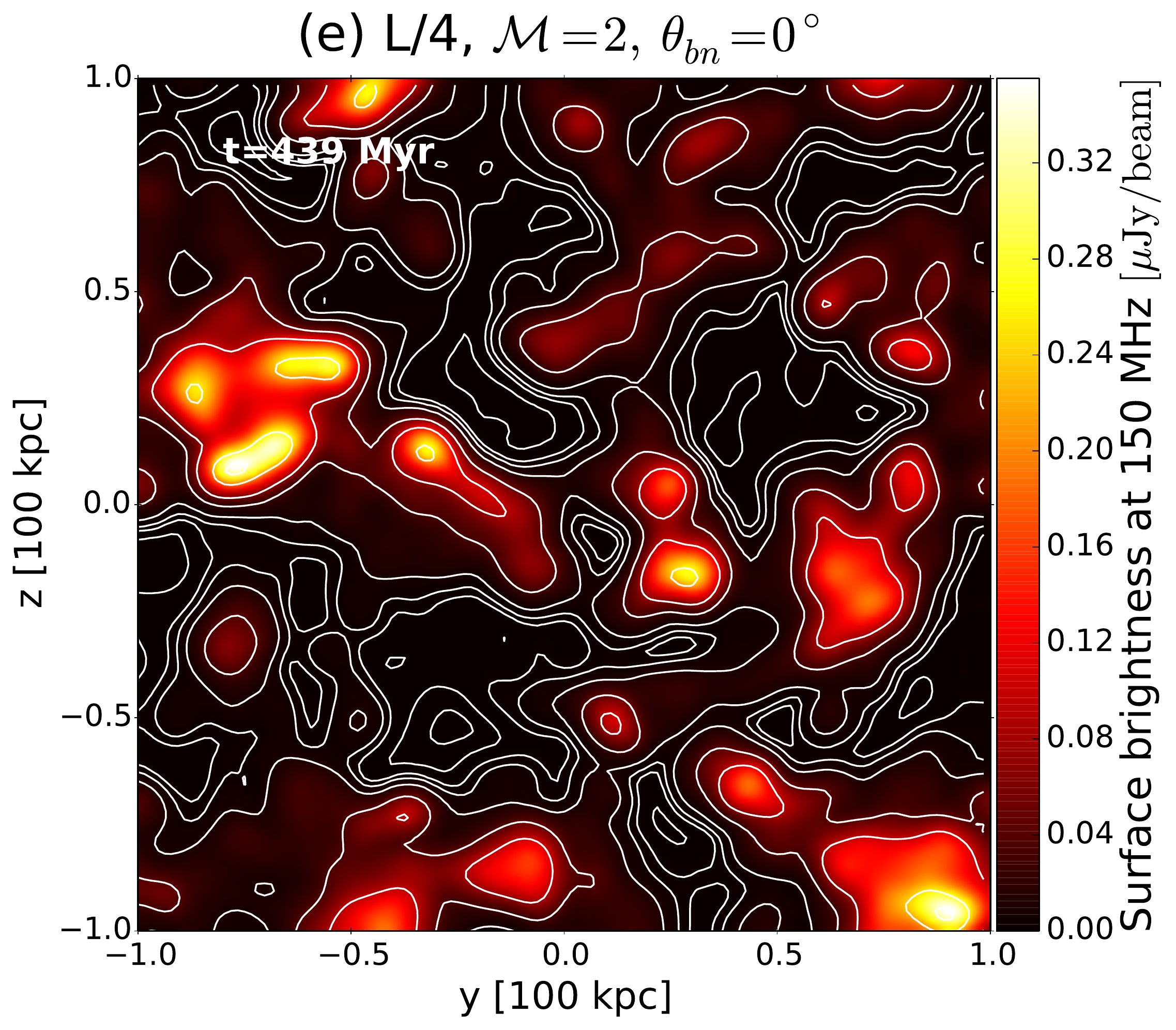}
  \includegraphics[width=.33\textwidth]{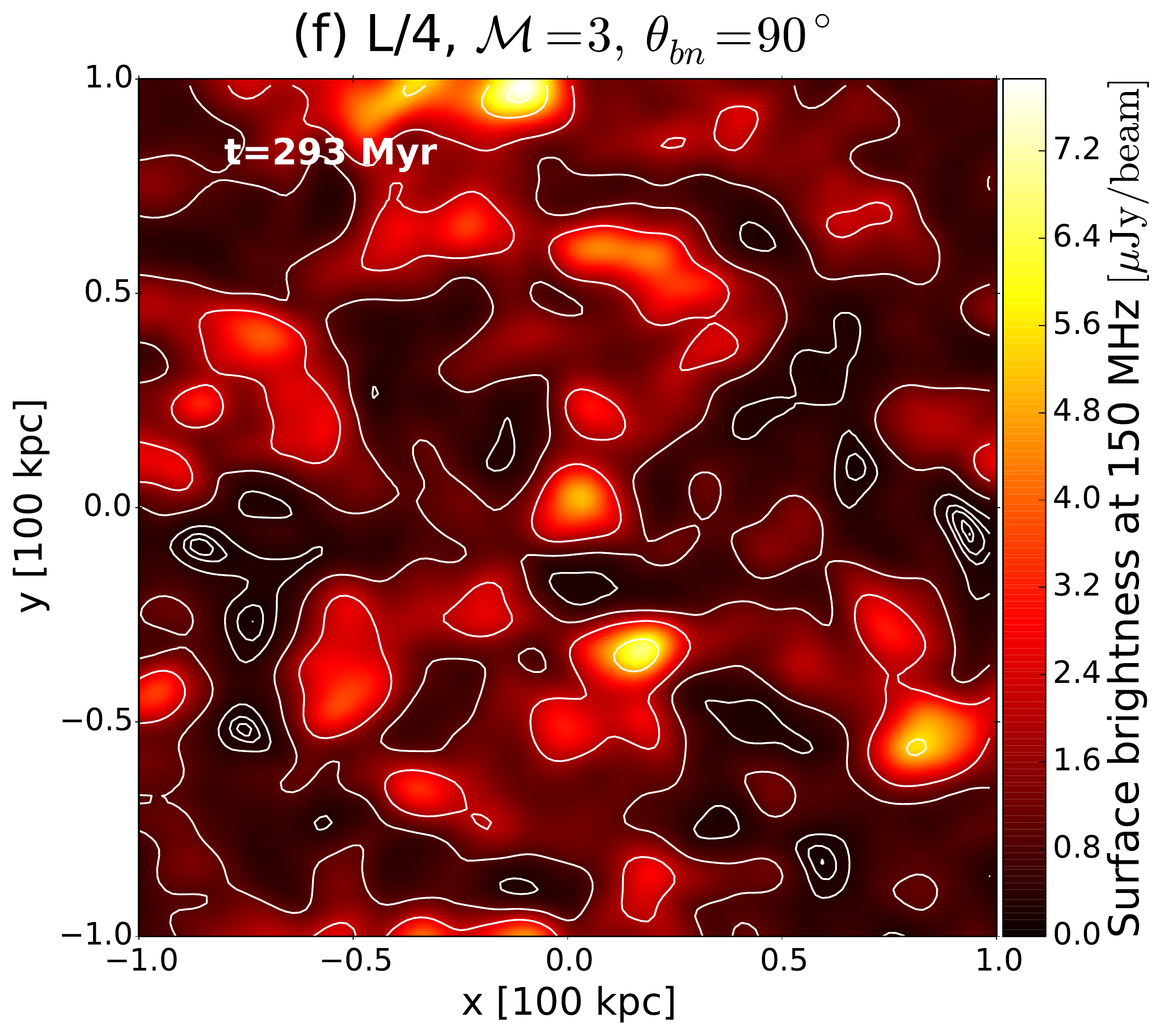}
    \caption{Surface brightness at 150 MHz integrated along the $x$-axis (in correspondence to Fig.~\ref{fig:emission_maps}). We considered a beam of $\theta^2 = 15"\times15"$ to get the surface brightness ($\theta^2 I_{\nu}$) in units of $\mu$Jy/beam. We smoothed the maps with a Gaussian kernel with $\mathrm{FWHM}=7.24$ kpc (assuming $z=0.023$).}
    \label{fig:emission_maps_X}
\end{figure*}

We present the surface brightness maps at 150 MHz for all our runs as viewed from the $x$-axis in Fig.~\ref{fig:emission_maps_X}. We want to point out that these maps were obtained by projecting the emissivity already used for this work along the $z$-axis. In order to compute self-consistently the surface brightness maps changing the LOS, one has to change the observing angle, $\theta_{obs}$, (see Sec.~\ref{sec:maps}) and the vector $\mathbf{\hat{n}_{los}}$. This is turn shall be used for computing the integral and Doppler factor in the emissivity equation (see Sec.~\ref{sec:sync_emiss}). The maximum value of surface brightness in each panel of Fig.~\ref{fig:emission_maps_X}, is lower than its correspondent panel in Fig.~\ref{fig:emission_maps}. This is because in this case, we end up summing up the contribution of the emissivity for a smaller region ($\sim 100$ kpc).

\end{document}